\documentclass[twoside,twocolumn,9pt]{article}

\newcounter{SoftMatter}
\usepackage{extsizes}
\usepackage[super,sort&compress,comma]{natbib}
\usepackage[version=3]{mhchem}
\usepackage[left=1.5cm, right=1.5cm, top=1.785cm, bottom=2.0cm]{geometry}
\usepackage{balance}
\usepackage{mathptmx}
\usepackage{amsmath}
\usepackage{amssymb}
\usepackage{xfrac}
\usepackage{sectsty}
\usepackage{graphicx}
\usepackage{lastpage}
\usepackage[format=plain,justification=justified,singlelinecheck=false,font={stretch=1.125,small,sf},labelfont=bf,labelsep=space]{caption}
\usepackage{float}
\usepackage{fancyhdr}
\usepackage{fnpos}
\usepackage{comment}
\usepackage{sidecap}
\usepackage{wrapfig}
\usepackage{subcaption}
\usepackage{tabularx}
\usepackage[english]{babel}
\addto{\captionsenglish}{%
  
}
\usepackage{array}
\usepackage{droidsans}
\usepackage{charter}
\usepackage[T1]{fontenc}
\usepackage[usenames,dvipsnames]{xcolor}
\usepackage{setspace}
\usepackage[compact]{titlesec}
\usepackage{hyperref}

\newcommand{\MM}[1]{{\color{cyan}MM: #1}}
\newcommand{\LW}[1]{{\color{magenta}LW: #1}}

\newcommand{\ie}{{\it i.e.}}

\newcommand{\rd}{{\rm d}}

\usepackage{epstopdf}

\begin{document}

\ifnum \value{SoftMatter}=1
\pagestyle{fancy}
\thispagestyle{plain}
\fancypagestyle{plain}{
\renewcommand{\headrulewidth}{0pt}
}

\makeFNbottom
\makeatletter
\renewcommand\LARGE{\@setfontsize\LARGE{15pt}{17}}
\renewcommand\Large{\@setfontsize\Large{12pt}{14}}
\renewcommand\large{\@setfontsize\large{10pt}{12}}
\renewcommand\footnotesize{\@setfontsize\footnotesize{7pt}{10}}
\makeatother

\renewcommand{\thefootnote}{\fnsymbol{footnote}}
\renewcommand\footnoterule{\vspace*{1pt}%
\color{cream}\hrule width 3.5in height 0.4pt \color{black}\vspace*{5pt}}
\setcounter{secnumdepth}{5}

\makeatletter
\renewcommand\@biblabel[1]{#1}
\renewcommand\@makefntext[1]%
{\noindent\makebox[0pt][r]{\@thefnmark\,}#1}
\makeatother
\renewcommand{\figurename}{\small{Fig.}~}
\sectionfont{\sffamily\Large}
\subsectionfont{\normalsize}
\subsubsectionfont{\bf}
\setstretch{1.125} 
\setlength{\skip\footins}{0.8cm}
\setlength{\footnotesep}{0.25cm}
\setlength{\jot}{10pt}
\titlespacing*{\section}{0pt}{4pt}{4pt}
\titlespacing*{\subsection}{0pt}{15pt}{1pt}

\fancyfoot{}
\fancyfoot[LO,RE]{\vspace{-7.1pt}\includegraphics[height=9pt]{head_foot/LF}}
\fancyfoot[CO]{\vspace{-7.1pt}\hspace{13.2cm}\includegraphics{head_foot/RF}}
\fancyfoot[CE]{\vspace{-7.2pt}\hspace{-14.2cm}\includegraphics{head_foot/RF}}
\fancyfoot[RO]{\footnotesize{\sffamily{1--\pageref{LastPage} ~\textbar  \hspace{2pt}\thepage}}}
\fancyfoot[LE]{\footnotesize{\sffamily{\thepage~\textbar\hspace{3.45cm} 1--\pageref{LastPage}}}}
\fancyhead{}
\renewcommand{\headrulewidth}{0pt}
\renewcommand{\footrulewidth}{0pt}
\setlength{\arrayrulewidth}{1pt}
\setlength{\columnsep}{6.5mm}
\setlength\bibsep{1pt}

\makeatletter
\newlength{\figrulesep}
\setlength{\figrulesep}{0.5\textfloatsep}

\newcommand{\topfigrule}{\vspace*{-1pt}%
\noindent{\color{cream}\rule[-\figrulesep]{\columnwidth}{1.5pt}} }

\newcommand{\botfigrule}{\vspace*{-2pt}%
\noindent{\color{cream}\rule[\figrulesep]{\columnwidth}{1.5pt}} }

\newcommand{\dblfigrule}{\vspace*{-1pt}%
\noindent{\color{cream}\rule[-\figrulesep]{\textwidth}{1.5pt}} }

\makeatother

\twocolumn[
  \begin{@twocolumnfalse}
{\includegraphics[height=30pt]{head_foot/SM}\hfill\raisebox{0pt}[0pt][0pt]{\includegraphics[height=55pt]{head_foot/RSC_LOGO_CMYK}}\\[1ex]
\includegraphics[width=18.5cm]{head_foot/header_bar}}\par
\vspace{1em}
\sffamily
\begin{tabular}{m{4.5cm} p{13.5cm} }

\includegraphics{head_foot/DOI} & \noindent\LARGE{\textbf{Switchable Wetting of Stimulus Responsive Polymer Brushes by Lipid Vesicles: Experiments and Simulations$^\dag$}} \\
\vspace{0.3cm} & \vspace{0.3cm} \\

 & \noindent\large{Felix Weissenfeld,$^{\ddag}$\textit{$^{a}$} Lucia Wesenberg,$^{\ddag}$\textit{$^{b}$} Masaki Nakahata, \textit{$^{c}$}  Marcus Müller,$^{\ast}$\textit{$^{b}$} and Motomu Tanaka$^{\ast}$\textit{$^{a,d}$}} \\

\includegraphics{head_foot/dates} & \noindent\normalsize{The abstract should be a single paragraph which summarises the content of the article. Any references in the abstract should be written out in full \textit{e.g.}\ [Surname \textit{et al., Journal Title}, 2000, \textbf{35}, 3523].} \\

\end{tabular}

 \end{@twocolumnfalse} \vspace{0.6cm}

  ]

\renewcommand*\rmdefault{bch}\normalfont\upshape
\rmfamily
\section*{}
\vspace{-1cm}


\footnotetext{\textit{$^{a}$~Physical Chemistry of Biosystems, Institute of Physical Chemistry, Heidelberg University, 69120 Heidelberg, Germany. E-mail: tanaka@uni-heidelberg.de}}
\footnotetext{\textit{$^{b}$~Institute for Theoretical Physics, Georg-August University, Friedrich-Hund-Platz 1, 37077 Göttingen, Germany. Email: mmueller@theorie.physik.uni-goettingen.de}}
\footnotetext{\textit{$^{c}$~Department of Materials Engineering Science, Graduate School of Engineering Science, Osaka University, 560-8531 Osaka, Japan}}
\footnotetext{\textit{$^{d}$~Center for Advanced Study, Institute for Advanced Study, Kyoto University, 606-8501 Kyoto, Japan}}
\footnotetext{\ddag~These authors contributed equally.}

\footnotetext{\dag~Electronic Supplementary Information (ESI) available: [details of any supplementary information available should be included here]. See DOI: 10.1039/cXsm00000x/}


\else

\title{\Large Switchable Wetting of Stimulus Responsive Polymer Brushes by Lipid Vesicles:\\Experiments and Simulations}
\author{Felix Weissenfeld,$^{\ddag}$ $^a$ Lucia Wesenberg,$^{\ddag}$ $^b$  Masaki Nakahata,$^c$ Marcus M{\"u}ller,$^{b,*}$  and Motomu Tanaka$^{a,d,*}$ \\
{\small\textit{$^{a}$~Physical Chemistry of Biosystems, Institute of Physical Chemistry, Heidelberg University, 69120 Heidelberg, Germany. E-mail: tanaka@uni-heidelberg.de}}\\
{\small\textit{$^{b}$~Institute for Theoretical Physics, Georg-August University, Friedrich-Hund-Platz 1, 37077 Göttingen, Germany. Email: mmueller@theorie.physik.uni-goettingen.de}}\\
{\small\textit{$^{c}$~Department of Materials Engineering Science, Graduate School of Engineering Science, Osaka University, 560-8531 Osaka, Japan}}\\
{\small \textit{$^{d}$~Center for Advanced Study, Institute for Advanced Study, Kyoto University, 606-8501 Kyoto, Japan}}\\
\ddag~These authors contributed equally.
}

\maketitle
\begin{abstract}
 We grafted polyacrylic acid brushes containing cysteine side chains at a defined surface density on planar lipid membranes. Specular X-ray reflectivity data indicated that the addition of Cd$^{2+}$ ions induces the compaction of the polymer brush layer and modulates the adhesion of lipid vesicles. The critical threshold level inducing the switch from non-wetting to partial wetting state, [Cd$^{2+}$]  = 0.25 mM, was determined by microinterferometry. The interactions between vesicles and brushes were quantitatively evaluated by height fluctuations of the membrane in contact with brushes and the shape of vesicles near the surface. To analyze these experiments, we have systematically studied adhesion of axially symmetric vesicles for finite-range membrane-substrate interaction and buoyancy, \ie, relevant experimental characteristics, through simulations. We found that (i) the local transversality condition that relates the maximal curvature at the edge of the adhesion zone to the adhesion strength remains rather accurate. Thus, although it is not experimentally possible to prepare vesicles at zero buoyancy, we can use the transversality condition to estimate the adhesion strength. We observe, however, that (ii) the adhesion diagram is significantly modified by a finite range of membrane-substrate interaction and buoyancy. For downward buoyancy, vesicles merely sediment onto the substrate and there is no mean-field adhesion transition. For upward buoyancy, adhered vesicles are metastable at best. Thus, a mean-field adhesion transition can only occur at vanishing buoyancy. Only for zero-range membrane-substrate interaction does a second-order adhesion transition occur at finite interaction strength. 
 For any finite-range interaction, the transition occurs when the membrane-substrate interaction changes from repulsive to attractive. We present a adhesion diagram as a function of adhesion strength and buoyancy and compare the adhesion behavior of vesicles to the wetting behavior of droplets of liquids.  
\end{abstract}

\fi
\section{Introduction}
Physical contact of cells to their neighbors, cell adhesion, plays key roles in a wide variety of biological processes. Cell adhesion modulates a number of biochemical signaling pathways \cite{Juliano2002Apr,Sackmann2021Feb} and tissue morphogenesis driven by forces acting between the neighboring cells.\cite{Heisenberg2013May} On the other hand, the inpaired cell adhesion functions are often associated with diseases, such as cancer metastasis. The significant reduction of cell-cell and cell-matrix adhesion causes the invasive migration and release of cancer cells into blood circulation.\cite{Cavallaro2004Feb} Ample evidence has indicated that cell adhesion is not only the static attachment between cells but also highly dynamic. For example, the freshwater polyp Hydra is able to regenerate the complete body with new head and foot by \textit{de novo} pattern formation from dissociated single cells by sorting cell-cell contacts.\cite{Gierer1972Sep,Technau1992May} On the molecular level, an increasing number of experimental studies have shown the dynamic rearrangement of adhesion molecules and their ligands play critical roles in immunological response \cite{Bromley2001Apr} and cell apoptosis. \cite{Balta2019Nov} Such experimental findings have been qualitatively recapitulated by using a phenomenological model of adhesion-induced phase separation \cite{Komura2000Nov}  or by assuming the presence of strong pinning centers.\cite{Frohlich2021Aug} However, the quantitative combination of experiments and simulations still remains challenging.  

Therefore, a large number of studies so far have been performed to physically model cell adhesion using rather simple, artificial lipid vesicles in the presence and absence of specific ligand-receptor like interaction pairs (stickers). Cell adhesion under equilibrium has been  described within the framework of wetting physics, irrespective of different origins on the molecular level. \cite{Bell1984Jun, Bruinsma2001Aug} In analogy to the shape of liquid drops on substrates, the shape of a cell or a lipid vesicle can be fine-tuned by switching the membrane-substrate interaction, $V(z)$, that quantifies the free energy of placing a unit area of membrane a distance, $z$, away from the synthetic substrate. $V(z)$ is characterized by the interplay of short- and long-range forces. Notably, what makes "biological" droplets, such as cells lipid vesicles, distinct from droplets of simple liquid is the interface between the interior and exterior; a bilayer lipid membrane. The intrinsic physical property of lipid bilayer membranes is their bending rigidity, $\kappa$, whereas the bilayer tension, $\gamma$, depends on the membrane geometry. This is in contrast to liquid drops where the interface tension of the liquid-vapor interface is an intrinsic property, independent from the drop shape.\cite{Helfrich1973Dec,Gompper1996Oct,Gueguen2017Sep} Another difference between vesicle adhesion and wetting of liquids is that the balance between adhesion energy and the free energy of the deformed vesicle or liquid drop does depend on the size, $R_0$, of the vesicle but remains invariant under scale changes for liquid drops. Thereby, one can systematically study the adhesion of vesicles by varying the vesicle size, leaving the surface chemistry unaltered.

From the experimental viewpoint, the use of soft polymer interlayers is a straightforward strategy to fine-adjust the adhesion of vesicles by tuning vesicle-substrate interaction, $V(z)$. Planar lipid membranes deposited on polymer supports, called polymer supported membranes, \cite{Tanaka2005Sep} have been used as soft “cushions” that reduce the frictional coupling of membranes and membrane-associated proteins by preventing the direct contact.\cite{Goennenwein2003Jul,Purrucker2007Feb} Previously, we measured specular neutron and X-ray reflectivity of zwitterionic phosphatidylcholine membranes deposited on about 20 and 40 nm-thick cellulose supports.\cite{Rossetti2015Apr} The equilibrium distance between membrane and underlying Si substrates determined by experiments could be reproduced by calculating the disjoining pressure including van-der-Waals, hydration repulsion, and Helfrich undulation quantitatively.

To switch the adhesion of vesicles, the use of stimulus responsive polymer brushes is a promising strategy, because they can change their physical properties (conformation, degrees of ionization, solvent affinity, etc.) by external cues, such as changes in temperature, pH, light, and ions. \cite{Alarcon2005Feb, Minko03b, Merlitz2009Jan, MWN, NatMat, Price2012Jan, Leonforte2016Dec, Brown2017Oct, Tanaka2020Aug} Previously, we transferred pH responsive diblock copolymers from the air/water interface to solid substrates and demonstrated the change in polymer chain conformation by pH titration. Intriguingly, the reversible change in polymer chain conformation led to a switching of the water layer between the membrane and brushes.\cite{Rehfeldt2006May} 

\begin{figure}[!tb]
\centering
  \includegraphics[width=0.45\textwidth]{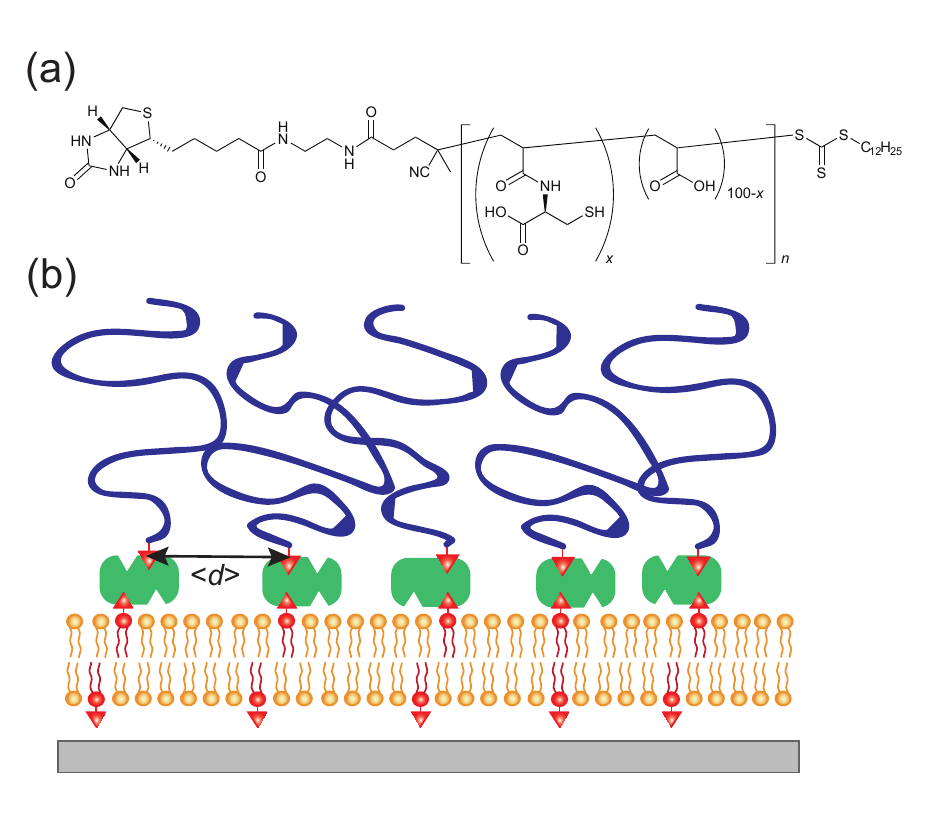}
  \caption{a) Chemical structure of PAA-Cys5-biotin. –COOH and –SH side chains act as the binding sites for Cd$^{2+}$. b) Grafting of PAA-Cys5-biotin brushes on supported membrane \textit{via} neutravidin crosslinkers.} 
  \label{figure_supported_membrane}
\end{figure}

In this study, we designed switchable polymer brush substrates that can switch $V(z)$ by forming chelator complexes with divalent ions in a concentration-dependent manner.  We synthesized polyacrylic acid brushes containing cysteine side chains terminated with biotin (PAA-Cys5-biotin, see \autoref{figure_supported_membrane}a) based on the hypothesis that --COOH and --SH side chains form a complex with Cd$^{2+}$ ions with a high affinity.\cite{Jalilehvand2011Nov} To achieve a uniform grafting of brushes at  a defined grafting density, we grafted the polymer chains onto supported membranes doped with biotin-functionalized lipids via neutravidin crosslinkers, instead of the commonly used "grafting onto" strategy. \cite{Kaindl2012Aug, Rieger2015Jan} Owing to the extremely high affinity of biotin and neutravidin, k$_D \approx 10^{-15}$ M,\cite{Helm1991} the average grafting distance, $\langle d \rangle$ can be controlled at nm accuracy simply by the doping ratio of biotin lipids (\autoref{figure_supported_membrane}b, see Method for more details). In contrast to previous studies inducing the change in surface charge density by the drastic change in solution pH \cite{Nardi1997Feb} or salt concentrations, \cite{Nardi1999Jun} PAA-Cys5 brushes enable to switch the conformation and hence $V(z)$ in the presence of 100 mM NaCl with 10 mM Tris buffer (pH 7.4), where the change in the total ionic strength and pH is negligible. The change in thickness, roughness and density of polymer brush layer was monitored by specular X-ray reflectivity, while the effective interfacial interaction potential was calculated from the height fluctuation of the membranes in contact with brushes. The global shape of vesicles (side view) was reconstructed from the confocal fluorescence microscopy images and compared to theory.

To this end, we represent the membrane by a thin elastic sheet within the Helfrich model \cite{Helfrich1973Dec} that has previously been utilized to study the adsorption of vesicles.\cite{Seifert1990Oct, Lipowsky1991Sep, Seifert1991Jun, Kraus1995Nov, Tordeux02b} Numerically minimizing the bending, adhesion, and potential energy in the gravitation field, we determine the optimal shape of the vesicle, paying particular attention to the effect of a finite range of the interaction, $V$, between membrane and substrate and buoyancy. Both effects are present in the experiment and have to be accounted for in a quantitative analysis. In particular, we devise an adhesion diagram for axially symmetric vesicles as a function of adhesion strength and buoyancy.

\section{Materials and Methods}
\subsection{Materials}
\subsubsection{Chemicals}
Milli-Q water from an ultrapurification system (Merck Millipore, Darmstadt, Germany) with a resistance > 18 M$\Omega$cm was used for all experiments. All the   1,2-dioleoyl-sn-glycero-3-phosphocholine (DOPC), 1,2-dioleoyl-sn-glycero-3-phosphoethanolamine-N-(cap biotinyl) (DOPE-biotin), Texas Red$^{TM}$ 1,2-Dihexadecanoyl-sn-Glycerin-3-Phosphoethanolamin (DHPE-Texas Red)  were purchased from Avanti Polar Lipids (Alabama, USA). Neutravidin was purchased from Thermo Fisher Scientific (Karlsruhe, Germany) and was ultracentrifuged before use (100.000 g, 1 h). Unless stated otherwise, the other chemicals were purchased from Sigma-Aldrich (Taufkirchen. Germany).
\subsubsection{Synthesis of PAA-Cys5-biotin}
PAA-Cys5-biotin (\autoref{figure_supported_membrane}a) was synthesized through copolymerization of \textit{S}-trityl-cysteine acrylamide (\textit{S}-Tri-Cys-AAm) and acrylic acid (AA) using 4,4'-((\textit{E})-diazene-1,2-diyl)bis(4-cyano-\textit{N}-(2-(5-((3a\textit{R},4\textit{R},6a\textit{S})-2-oxohexahydro-1\textit{H}-thieno[3,4-\textit{d}]imidazol-4-yl)pentanamido)ethyl)pentanamide) (ACVA-biotin) as an initiator and 2-(dodecylthiocarbonothioylthio)-2-methylpropionic acid (DDMAT) as a chain transfer agent, followed by deprotection of trityl group with trifluoroacetic acid (TFA). In brief: \textit{S}-Tri-Cys-AAm (0.05 mmol), AA (0.95 mmol), ACVA-biotin (0.01 mmol), and DDMAT (0.01 mmol) were dissolved in 1 mL of dimethylsulfoxide (DMSO) dried with molecular sieves 4A. The solution was purged with nitrogen gas for 1 h, sealed, and heated in an oil bath at 70 °C overnight. The solution was poured into acetone (10 mL) with stirring. The resultant oily precipitate was collected with centrifugation (3,500 rpm, 5 min.). After removing supernatant by decantation, trifluoroacetic acid (TFA) (1 mL) was added and stirred for overnight. The solution was poured into diethyl ether (10 mL). The resultant solid precipitate was washed with diethyl ether (10 mL) three times and dried in vacuum at room temperature. Successful polymerization and deprotection were confirmed by 1H NMR spectra recorded at 400 MHz with a JNM–ECS400 NMR spectrometer (JEOL, Tokyo, Japan). Gel permeation chromatography (GPC) analysis of PAA-Cys5-biotin was carried out using a GL-7400 HPLC system (GL Science, Tokyo, Japan) equipped with Inertsil WP300 Diol column (GL Science, Tokyo, Japan) and refractive index (RI) detector (RID-20A, Shimadzu, Kyoto, Japan), using PBS as an eluent at the flow rate of 0.3 mL/min. at 25 ºC. ReadyCal-Kit Pullulan (purchased from PSS Polymer Standards Service GmbH, Mainz, Germany) was used as the calibration standard. The weight average molecular weight (\textit{M}$_{w}$) and polydispersity index (\textit{M}$_w$/\textit{M}$_n$) of PAA-Cys5-biotin were estimated to be \textit{M}$_w$ = 7.4 × 10$^{4}$ Da and \textit{M}$_w$/\textit{M}$_n$ = 2.5, respectively.

\subsection{Methods}
\subsubsection{Preparation of vesicle suspensions}
Lipid stock solutions in chloroform (2 mg mL$^{-1}$) containing 98 mol\% DOPC and 2 mol\% DOPE-biotin were stored in a vacuum oven overnight.The dried lipid mixture was suspended in a Tris (10 mM) buffer containing 100 mM NaCl (pH 7.4). Small unilamellar vesicles were prepared by sonication of a lipid suspension with a Misonix Sonicator 3000 (Misonix, Düsseldorf, Germany). Giant unilamellar vesicles were prepared by electro-swelling as reported previously \cite{matsuzaki2017adsorption}. In brief: indium tin oxide (ITO) coated glass slides (Sigma-Aldrich) were spin-coated with DOPC including 0.2 mol\% DHPE-Texas Red. The dried lipid mixture was hydrated with sucrose solutions under AC potentials (10 Hz, 3V)  at 37 °C for 2 h. Finally a 0.2 mL portion of the vesicle suspension was mixed with 1 mL of the medium to adjust buoyancy and osmolality. The osmolality was measured by an micro-osmometer (OM 806 , Löser, Berlin, Germany). To enable the vesicles to adhere to the brush surface, the difference in the density of medium inside and outside of the vesicle was adjusted at different levels: $\Delta \rho_{1} = -20 $kg m$^{-3}$, $\Delta \rho_{2} = +20 $kg m$^{-3}$, $\Delta \rho_{3} = +40 $kg m$^{-3}$. It should be noted that the buoyancy strongly depends on the vesicle size $ R_0$:\cite{Kraus1995Nov}
\begin{equation}
\tilde \epsilon_B = \frac{\Delta \rho g R_0^4}{\kappa}.
\end{equation}

\subsubsection{Grafting of PAA-Cys5-biotin brushes on supported membranes}
PAA-Cys5-biotin brushes were grafted on the surface of DOPC membranes deposited on planar solid substrates (supported membranes) doped with 2 mol\% of DOPE-biotin. The major advantage of supported membranes over commonly used is the direct grafting of polymers onto solid substrates via covalent bonds (called “grafting onto” method) is their capability to achieve high and well defined grafting densities. As DOPC and DOPE-biotin are miscible, the average grafting distance <\textit{d}> can be controlled at nm accuracy simply by the doping ratio of DOPE-biotin lipids $\chi_{biotin},  <\textit{d}> =  (0.6/\chi_{biotin})^{0.5}$ [nm], assuming the area per lipid molecule is 0.6 nm$^2$ \cite{Lipowsky_Sackmann1995}. A supported membrane was deposited by incubating vesicle suspension with a cleaned glass slide \cite{Kern_Putoinen1983, Hillebrandt_Tanaka1983} for 30 min. at 40°C, and the unbound SUVs were carefully removed by rinsing. In the next step the sample was incubated with neutravidin (40 $\mu$g/mL) for 1 h at 40°C. After removing unbound neutravidin, the solution of PAA-Cys5-biotin (40 $\mu$g/mL) was injected and incubated under the same conditions. 

\subsubsection{Confocal fluorescence microscopy imaging of global vesicle shape}
The global shape of GUVs was captured with a Nikon A1R confocal microscope (Nikon Europe, Düsseldorf, Germany) 60 $\times$ oil immersion objective (NA 1.40). For the fluorescence imaging, GUVs were labeled by incorporating 0.2 mol\% of TexasRed-DHPE. The vesicles were deposited on brush surfaces pre equilibrated with [CdCl$_2$] = 0-1.0 mM for 30 min before the imaging. Confocal side view images were obtained due to confocal 3D reconstruction of confocal bottom view images with a stepsize of 0.5 µm using ImageJ. The distortion of the reconstructed image in z-direction was corrected by taking the image of commercially available fluorescently labeled latex particles with a similar size (\textit{R} = 7.5 $\mu$m).

\subsubsection{Specular X-ray reflectivity}
X-ray reflectivity curves were measured using a D8 Advance diffractometer (Bruker, Germany) operating with a sealed X-ray tube emitting
Mo K$_\alpha$ radiation (E = 17.48 keV, $\lambda$ = 0.0709 nm). The beam size was defined to 200 $\mu$m in the scattering plane after its collimation by various slits. To avoid the sample radiation damage, the attenuator was set to automatic. The cleaned Si wafers were placed into a Teflon chamber with Kapton windows, and the momentum transfer normal to the plane of the membrane is given as a function of the angle of incidence $\alpha_{i}$, 
\begin{equation}
q_z = \frac{4\pi}{\lambda}\sin \alpha_i.
\end{equation}

The reflectivity for each data point was corrected for the beam footprint and for the beam intensity. A genetic minimization algorithm of the Parratt formalism \cite{Parratt1954Jul} implemented in the Motofit software \cite{Nelson2006Apr} was used to fit the experimental data.

\subsubsection{Label-free, microinterferometry imaging of vesicle/brush contact}
The interaction between vesicles and brush substrates were monitored by label-free, reflection interference contrast microscopy (RICM) \cite{albersdorfer1997adhesion,limozin2009quantitative,Frohlich2021Aug} RICM imaging was performed on an Axio Observer Z1 microscope (Zeiss, Oberkochen, Germany) equipped with a 63$\times$ oil immersion objective (NA 1.25) with a built in $\lambda/4$ plate. To record multiple interferences, the Illumination Numerical Aperture (INA) was adjusted to 0.48, about 400 consecutive images were collected by an Orca-Flash4.0LT camera (Hamamatsu Photonics, Herrsching, Germany) at an exposure time of 30 ms, and were subjected to the analysis. The intensity I was converted to the relative height $\delta z$ via: 
\begin{equation}
I(\delta z)=I_{1}+I_{2} + 2 \sqrt{I_{1}I_{2}} \cos(2k\delta z(x,y)+\Phi).
\end{equation}

$I_i$ represents the intensity of the light reflected at the \textit{i}th interface,  and $k = \frac{2\pi n}{\lambda}$ is the wave vector. n is the refractive index of the buffer (n = 1.344), $\lambda$ the wavelength ($\lambda=475$ nm), and $\Phi$ the phase shift of the reflected light. To monitor the height fluctuation h(t), we collected the mean intensity from 3 $\times$ 3 pixels as a function of time I(t):
\begin{equation}
    \delta z(t) = \arccos \left(\frac{2I(t)-(I_{\max}+I_{\min})}{(I_{\max}-I_{\min})} \right)
    \frac{\lambda}{4\pi n}.
    \label{eqn:RICM_heigth}
\end{equation}
$I_{\max}$ and $I_{\min}$ are the intensity minima and maxima of the background. The analyses were performed using a self-written Matlab routine (R2019a) \cite{Frohlich2021Aug}.

\subsection{Model and numerical techniques}

\subsubsection{Description of vesicle shape}

\begin{figure}[h]
\centering
  \includegraphics[width=\columnwidth]{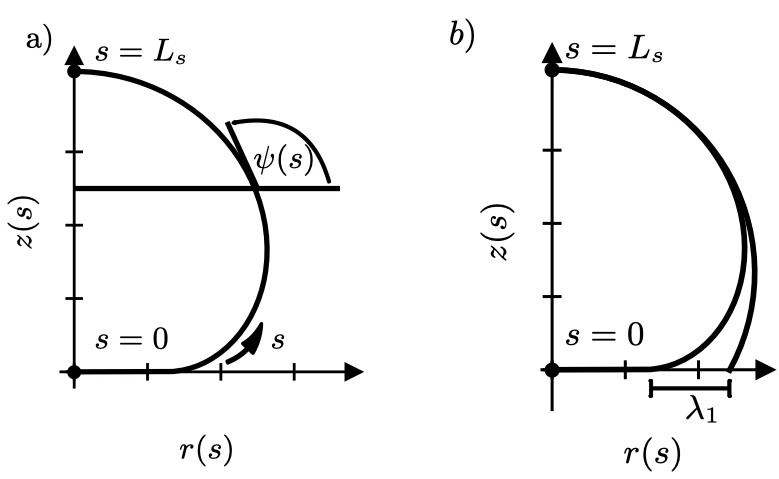}
  \caption{
  a) Parameterization of an axially symmetric vesicle. The arc-length parameter, $0\leq s < L_s$, runs from the bottom center, $z(s=0)=z_0$, $r(s=0)=0$, to the top $z(s=L_s)=H$, $r(s=Ls)=0$.  The angle, $\psi(s)$, denotes the angle between the tangent of the vesicles contour and the horizontal.
  b) Comparison between the vesicle shape, parameterized by $\psi(s),z_0,L_s$, and a spherical cap, characterized by the wetting angle, $\Theta$, and its radius $R_{\rm cs}$.  The length, $\lambda_1$, is the distance between the contact radius of the spherical cap and that of the vesicle. 
  }
  \label{fgr:model}
\end{figure}

We limit our considerations to axially symmetric vesicle using a cylindrical coordinates, $(z,r)$. Deviations from axial symmetry in the presence of buoyancy have been considered in Ref.~\cite{Kraus1995Nov}. The assumption of axial symmetry, however, simplifies the analysis and is compatible with the experimental observations in our work.  The membrane area $A_0=4\pi R_0^2$ is fixed, but the enclosed volume is unconstraint, \ie, the membrane is assumed to be permeable on the experimental time scale. The vesicle interacts with a solid substrate via a potential $V_w(z)$, where $z$ denotes the distance from the substrate. $z=0$ denotes the position of the minimum of $V_w(z)$. As illustrated in \autoref{fgr:model} we parameterize the vesicle shape by the tangent angle, $\psi(s)$, as a function of the contour length, $0\leq s < L_s$.\cite{Lipowsky1991Sep,Tordeux02b,Gozdz2004Aug}

For the numerical minimization of the vesicle (free) energy, we expand $\psi(s)$ around a spherical vesicle in a Fourier series \cite{Gozdz2004Aug, Raval2020Jul}
\begin{equation}
\psi(s) = \frac{\pi}{L_s}s + \sum_{k=0}^{N_{\rm max}} a_k \sin\left(\frac{\pi[k+1]}{L_s}s \right)
\label{eqn:psi}
\end{equation}
In the absence of a substrate or volume interactions, the vesicle shape is spherical. Its radius, $R_0$, is set by the membrane area, $L_s=\pi R_0$, and $a_k=0$ for all $k=0,\cdots,N_{\rm max}-1$. This parameterization fulfills the boundary condition, $\psi(0)=0$ and $\psi(L_s)=\pi$. The arc length, $L_s$, must be chosen such that $r(L_s)=0$. The position of the axially symmetric vesicle membrane takes the form
\begin{eqnarray}
r(s) &=& \int_0^s \rd s'\; \cos(\psi(s')) \label{eqn:r} \\
z(s) &=& z_0+\int_0^s \rd s'\; \sin(\psi(s')) \label{eqn:z}
\end{eqnarray}
where the position, $z_0$, at the bottom center specifies the position of the vesicle along the symmetry axis. Thus, the vesicle shape and position are specified by $z_0$, $L_s$, and the Fourier coefficients $a_k$ with $k=0,\cdots,N_{\rm max}-1$ of $\psi(s)$, obeying the constraint $r(L_s)=0$.

\subsubsection{Energy of the vesicle -- bending, adhesion, and buoyancy}

The energy of the vesicle is comprised of three contributions: bending energy,\cite{Helfrich1973Dec} interaction with the substrate, \cite{Seifert1990Oct, Lipowsky1991Sep, Seifert1991Jun, Tordeux02b} and buoyancy.\cite{Kraus1995Nov} 

We represent the bending energy by the Helfrich Hamiltonian, ${\cal H}_b$, that expresses the energy costs via a surface integral over the two principle curvatures, $C_1$ and $C_2$. \cite{Helfrich1973} Using the parameterization, $\psi(s)$, these curvatures take the form \cite{Lipowsky1991Sep, Tordeux02b, Gozdz2004Aug}
\begin{eqnarray}
C_1(s) = \frac{\rd \psi}{\rd s} &=&  \frac{\pi}{L_s} + \sum_{k=0}^{N_{\rm max}} a_k \frac{\pi[k+1]}{L_s} \cos\left(\frac{\pi[k+1]}{L_s}s \right)
\label{eqn;C1} \\
C_2(s) &=& \frac{\sin(\psi(s))}{r(s)} 
\label{eqn;C2}
\end{eqnarray}
Since the two membrane leaflets are symmetric, and the spontaneous curvature of the membrane vanishes. Likewise we have assumed that the membrane is homogeneous such that the Gaussian curvature term only provides a constant contribution and needs not to be considered. Thus the bending energy takes the simple form \cite{Lipowsky1991Sep, Tordeux02b, Gozdz2004Aug}
\begin{equation}
{\cal H}_{b} = \frac{\kappa}{2}  \int _0^{L_s} \rd s\; 2 \pi r \; \left( \frac{\rd \psi}{\rd s} + \frac{\sin(\psi)}{r}  \right)^2
\label{eqn:Helfrich}
\end{equation}
The parameter, $\kappa$, denotes the bending energy of the membrane, and it sets the energy scale. The integration of ${\cal H}_b$ and subsequent quantities is performed numerically in the $zr$-plane by discretizing the parametric vesicle shape, $r(s)$, $z(s)$ into $N_s$ points. Using a trapezoidal integration, the error is on the order $N_s^{-2}$.

\begin{figure}[!tbh]
\centering
  \includegraphics[width=0.5\textwidth]{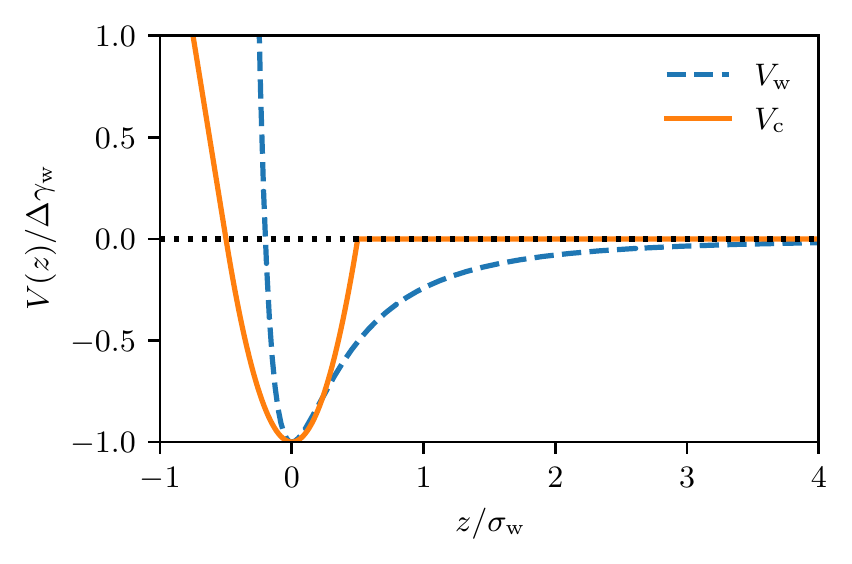}
  \caption{Illustration of the long-range and short-range membrane-substrate interactions, $V_w(z)$ and $V_c(z)$, according to \autoref{eqn:Vw} and \autoref{eqn:Vc}. }
  \label{fig:1}
\end{figure}

Whereas prior studies often modeled the interaction between vesicle membrane and substrate per unit area by a contact potential \cite{Seifert1990Oct, Lipowsky1991Sep, Tordeux02b} (see Ref.~\cite{Seifert1991Jun} for an exception), we consider short-range potentials, $V_c(z)$, with a finite width, $\sigma_w$, and long-range potentials, $V_w(z)$, that represent van-der-Waals interactions. The two types of potentials are illustrated in \autoref{fig:1}. The origin of the $z$-axis is the minimum of the membrane-substrate potential, and $-\Delta \gamma_w$ denotes the value of the membrane-substrate potential at its minimum. The short-range potential takes the form
\begin{equation}
V_c(z) =  \left\{ \begin{array}{ll} 
- 4 \Delta \gamma_w \left(\frac{z}{\sigma_w}+\frac{1}{2}\right) & \mbox{for}\quad z< -\frac{\sigma_w}{2} \\
- \Delta \gamma_w \left(1-\frac{2z}{\sigma_w}\right) \left(1+\frac{2z}{\sigma_w}\right) & \mbox{for}\quad |z|<\frac{\sigma_w}{2} \\
  0 & \mbox{otherwise}
  \end{array} \right.
\label{eqn:Vc}
\end{equation}
and long-range interactions take the Hamaker-form
\begin{equation}
V_w(z) =  \frac{3\sqrt{3}}{2}  \Delta \gamma_w\left( {\rm sign}( \Delta \gamma_w)\left[\frac{\sigma_w}{z+z_w}\right]^9- \left[\frac{\sigma_w}{z+z_w}\right]^3 \right)
\label{eqn:Vw}
\end{equation}
where the sign-function assures that potential remains repulsive even for $\Delta \gamma_w<0$. $z_w=\sqrt[6]{3}\sigma_w$ shifts the minimum of $V_w(z)$ to $z=0$. We note that long-range power-law decay is scale-free. 

Integrating the membrane-substrate interaction over the vesicle, we obtain the adhesion energy
\begin{equation}
{\cal H}_{w} =  \int _0^{L_s} \rd s\; 2 \pi r(s) \; V(z(s))
\label{eqn:Hw}
\end{equation}

Instead of characterizing the range of the potential by $\sigma_w$, we can use the curvature of the potential at its minimum. The latter quantity controls the experimentally accessible, thermal height fluctuations of a membrane bound to the substrate. For the two types of potential we obtain
\begin{eqnarray}
\left.\frac{\rd^2 V_c}{\rd z^2}\right|_0 &=& - 8\frac{\Delta \gamma_w}{\sigma_w^2} \\
\left.\frac{\rd^2 V_w}{\rd z^2}\right|_0 &=& 9\sqrt[6]{81}\frac{\Delta \gamma_w}{\sigma_w^2}
\end{eqnarray}

Finally, experiments often employ buoyancy to bring the vesicles to the surface. In this case, there is a mass-density difference, $\Delta \rho$, between the liquid enclosed by the vesicle and the surrounding solution. The potential energy in the gravitational field takes the form
\begin{equation}
{\cal H}_{g} =  \Delta \rho g \int _0^{L_s} \underbrace{\rd s\; \sin(\psi(s))}_{=\rd z} \; \pi r^2(s) \;  z
\end{equation}
where $g$ denotes the gravitational acceleration constant. Upward buoyancy corresponds to $\Delta \rho g<0$.

The total energy is given by the sum of these three contribution, ${\cal H}_0 = {\cal H}_{b} + {\cal H}_{w} + {\cal H}_{g}$.

\subsubsection{Restraints}

In the following we seek to minimize the vesicle energy, ${\cal H}_0[\psi,z_0,L_s]$, under the following constraints: (i) fixed membrane area, $A[\psi]=4\pi R_0^2$ and (ii) vesicle closure, $r(L_s)=0$. Additionally, we could enforce (iii) the position of the $z$-coordinate of the enclosed volume, $Z=z_{\rm cm}$, or (iv) the volume, $V_0$, enclosed by the vesicle. Numerically, the constraints are mollified, and the resulting restraints are incorporated into the energy functional via umbrella potentials with large spring constants. 

Deviations of the membrane area, $A[\psi,L_s]$, from the reference value, $A_0 = 4 \pi R_0^2$, increase the energy by an amount
\begin{equation}
H_A = \frac{k_A}{2} \left( A[\psi,L_s]-4\pi R_0^2\right)^2
\end{equation}
with
\begin{equation}
A[\psi,L_s] =  \int_0^{L_s}\rd s\; 2\pi r(s)
\label{eqn:A}
\end{equation}
From the deviations we can estimate the membrane tension
\begin{equation}
\gamma = k_A  \left( A[\psi,L_s]-4\pi R_0^2\right)
\end{equation}
in the large $k_A$-limit, \ie, $k_A$ is related to the inverse areal compressibility of the membrane.

Likewise, vesicle closure, $r(L_s)=0$, gives rise to the contribution
\begin{equation}
H_r = \frac{k_r}{2}  \left(r(L_s)\right)^2
\end{equation}

To make connection to wetting transition it is instructive to restrain the vesicle's center-of-mass $Z[\psi,z_0,L_s] = \frac{{\cal H}_{g} }{ \Delta \rho g \; V}$ \textit{via}
\begin{equation}
H_p = \frac{k_p}{2}  \left(Z[\psi,z_0,L_s]-z_{\rm cm}\right)^2 \label{eqn:rzcm}
\end{equation}

For completeness, we mention that deviations of the enclosed volume $V[\psi,L_s] = \int_0^{L_s}\rd s\; \sin(\psi) \; \pi r^2$ from a reference value, $V_0$, could be penalized by an energy contribution
\begin{equation}
H_V = \frac{k_V}{2}  \left(V[\psi,L_s]-V_0)\right)^2
\end{equation}
yielding the pressure difference $\Delta P = k_V  \left(V[\psi,L_s]-V_0)\right)$ across the membrane for $k_V \to \infty$. In the following, however, we set $k_V=0$ and let the vesicle's volume adjust.

The total energy, ${\cal H}={\cal H}_0 + \delta H$, to be minimized contains the three energies, ${\cal H}_0 = {\cal H}_{b} + {\cal H}_{w} + {\cal H}_{g}$, and restraints, $\delta H= H_A+H_V+H_r+H_p$. In the following we measure all energies in units of the membrane's bending rigidity, $\kappa$, and all length scales in units of the radius, $R_0$, of a spherical vesicle with the same membrane area as the restraint, $A_0$. 
\begin{eqnarray}
\frac{{\cal H}[\psi,z_0,L_s]}{\kappa} &=&  \pi \int _0^{L_s} \frac{\rd s}{R_0}\;  \frac{r}{R_0} \; \left( R_0 \frac{\rd \psi}{\rd s} + R_0\frac{\sin(\psi)}{r}  \right)^2 
\label{eqn:Hall}\\
&& +  \underbrace{\frac{\Delta \gamma_w R_0^2}{\kappa}}_{\equiv \tilde \epsilon_w} \int _0^{L_s} \frac{\rd s}{R_0}\;  \frac{2\pi r}{R_0} \; \frac{V_w(z)}{\Delta \gamma_w} \nonumber \\
&& + \underbrace{\frac{\Delta \rho g R_0^4}{\kappa}}_{\equiv -\tilde \epsilon_B} \int _0^{L_s} \frac{\rd s}{R_0}\; \sin(\psi) \;  \frac{\pi r^2  z}{R_0^3} \nonumber \\
&& +\frac{k_A R_0^4}{2\kappa} \left( \frac{A[\psi,L_s]}{R_0^2}-4\pi\right)^2 \nonumber \\
&& +\frac{k_V R_0^6}{2\kappa}  \left(\frac{V[\psi,L_s]-V_0}{R_0^3})\right)^2 \nonumber \\
&& +\frac{k_rR_0^2}{2\kappa}  \left(\frac{r[\psi,L_s](L_s)}{R_0}\right)^2 \nonumber \\
&& +\frac{k_p R_0^2}{2\kappa}  \left(\frac{Z[\psi,z_0,L_s]-z_{\rm cm}}{R_0}\right)^2 \nonumber
\end{eqnarray}
The thermodynamic state of the vesicle in contact with a solid substrate is characterized by two dimensionless parameter combinations, $\tilde\epsilon_w=\frac{\Delta \gamma_w R_0^2}{\kappa}$ \cite{Seifert1990Oct, Lipowsky1991Sep} and $\tilde\epsilon_B=\frac{\Delta \rho g R_0^4}{\kappa}$ \cite{Kraus1995Nov} that measure the relative strength of adhesion and buoyancy with respect to the bending energy, respectively. 

The energy functional, ${\cal H}$ is numerical minimized with respect to $\psi,z_0,L_s$ by a conjugate-gradient method. Typical parameter values are compiled in \autoref{tab:1}. The values that minimize ${\cal H}[\psi,z_0,L_s]$ are denoted by $\psi^*,L_s^*,z_0^*$; these values depend on the thermodynamic state, specified by $\tilde \epsilon_w$ and $\epsilon_B$. Inserting these values into the energy functional, we obtain the energy, $H={\cal H}[\psi^*,L_s^*,z_0^*]$ of the vesicle at a given thermodynamic state. Note that this procedure completely ignores thermal fluctuations.

\begin{table}
\centering
\begin{tabularx}{0.5\textwidth}{ll}
\hline \hline
variable & value \\
\hline
$N_{\rm max}$ & $128,144$  \\
$N_s$ & $578, 2048$ \\
$\sigma_w/R_0$ & $0.03, 0.01$ or $0.002$  \\
$\tilde k_A\equiv k_AR_0^4/\kappa$ & $100$  \\
$\tilde k_V\equiv k_VR_0^6/\kappa$ & $0$, permeable membrane\\
$\tilde k_r\equiv k_rR_0^2/\kappa$ & $1000$  \\
$\tilde k_p\equiv k_pR_0^2/\kappa$ & $10000$ or $0$  \\
\hline \hline
\end{tabularx}
\caption{\label{tab:1} Compilation of typical parameters for the numerical minimization of the vesicle shape.
}
\end{table}

To study the adhesion transition we quantify the dimensionless energy difference between the vesicles in contact with a substrate and a free, unbound vesicle in the absence of buoyancy, $H=H_b=8\pi\kappa$
\begin{equation}
\tilde f=\frac{{\cal H}_0[\psi^*,L_s^*,z_0^*]}{8\pi \kappa}-1
\end{equation}
\section{Results}

\subsection{Switching of brush conformation by chemical stimulus}

\begin{figure}[!tbhp]
\centering
  \includegraphics[width=0.4\textwidth]{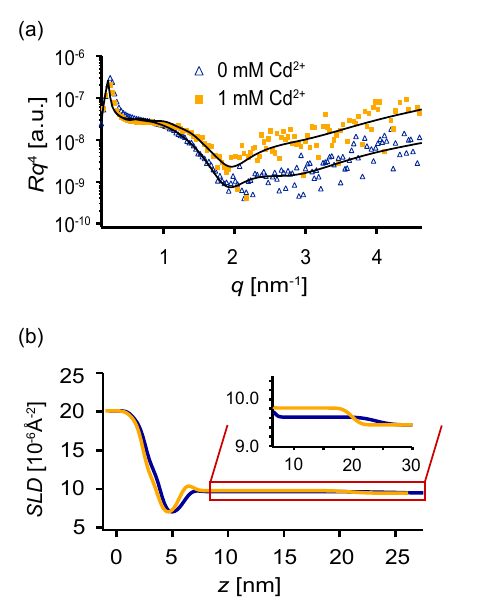}
  \caption{a) High energy specular X-ray reflectivity curves measured at [Cd$^{2+}$] = 0 mM (blue) and 1 mM (yellow). The best fit results are presented by solid lines. b) Scattering length density (SLD) profiles reconstructed from the best fit results. The magnified profiles near the interface are shown in inset.}
  \label{figure_XRR_SLD}
\end{figure}

\autoref{figure_XRR_SLD}a and \autoref{figure_XRR_SLD}b  show the X-ray reflectivity data and the scattering length density profiles reconstructed from the fitting. The addition of 1 mM Cd$^{2+}$ ions caused a clear change in the global shape of the X-ray reflectivity, suggesting that PAA-Cys5-biotin brushes change their conformation. The best fit results \autoref{tab:M1} and the magnified view of the scattering length density in the vicinity of interface suggest that the brush layer thickness decreased from $d_0$ = 1.43 nm to $d_{\rm Cd}$ = 1.03  nm, accompanied by the decrease in the brush/solution interface roughness from $\sigma_0$ = 2.0 nm  to $\sigma_{\rm Cd}$ = 1.2 nm. The obtained results indicate that the addition of 1 mM Cd$^{2+}$ ions makes PAA-Cys5-biotin brush more compact and the density gradient from the brush regime to the bulk solution sharper. The compaction of PAA-Cys5-biotin caused by additional Cd$^{2+}$ ions might be attributed to the screening of intra- and interchain electrostatic repulsions between negatively charged –COOH side chains by capturing of positively charged Cd$^{2+}$ ions. 
In fact, the Zeta potential of PAA-Cys5-biotin brushes grafted on supported membranes deposited on SiO$_2$ microparticles (radius: 1.5 $\mu$m) showed a monotonic increase from $\zeta_{1\mu {\rm M}}$ = – 26 mV to $\zeta_{1 {\rm mM}}$ = – 10 mV (see Appendix). 
This qualitatively agrees well the previous report, showing that the decrease in the ionization degree of side chains results in the compaction of polyelectrolyte brushes \cite{Rehfeldt2006May}. 
Nevertheless, the addition of 1 mM Cd$^{2+}$ ions does not not alter pH (7.4) or Debye screening length ($\kappa_{\rm D}^{-1}$ < 10 {\AA}) because the solution contains 100 mM NaCl buffered with 10 mM Tris (pH 7.4). The compaction of PAA-Cys5-biotin brushes in the presence of divalent Cd$^{2+}$ ions might be attributed to the complex formation by Cd$^{2+}$ ion with --COOH and --SH side chains \cite{Jalilehvand2011Nov}. 
Further spectroscopic studies under the systematic variation of the fraction of cystein side chains will help us understand the molecular-level mechanism.

\begin{table}
\caption{\label{tab:M1} Thickness $d$, scattering length density $SLD$ and roughness $\sigma$ corresponding to best fit results of XRR data \autoref{figure_XRR_SLD}.}
\centering
\begin{tabular}{llll}
\hline \hline
\multicolumn{4}{c}{PAA-Cys5-biotin in the absence of Cd$^{2+}$ } \\
\hline
 & $d$  & $SLD$ & $\sigma$ \\
 & (nm) & (10$^{-6}$\AA$^{-2}$) & (nm) \\
\hline
SiO$_2$ & 1.23 $\pm$ 0.03 & 18.9 & 0.55 $\pm$ 0.01 \\
buffer & 0.46 $\pm$ 0.01 & 9.45 & 0.45 $\pm$ 0.02 \\
lipid headgroup$_{inner}$ & 0.68 $\pm$ 0.01 & 13.1 $\pm$ 0.2 & 0.47 $\pm$ 0.01  \\
lipid alkylchain & 2.24 $\pm$ 0.03 & 6.8 $\pm$ 0.01 & 0.59 $\pm$ 0.02 \\
lipid headgroup$_{outer}$ & 0.89 $\pm$ 0.05 & 12.3 $\pm$ 0.2 & 0.62 $\pm$ 0.06\\
neutravidin +  & 18.2 $\pm$ 0.8 & 9.7 $\pm$ 0.1 & 2.05 $\pm$ 0.1\\
PAA-Cys5-biotin & & & \\
\hline \hline
\multicolumn{4}{c}{PAA-Cys5-biotin in the presence of 1mM Cd$^{2+}$} \\
\hline
 & $d$  & $SLD$ & $\sigma$ \\
 & (nm) & (10$^{-6}$\AA$^{-2}$) & (nm) \\
\hline 
SiO$_2$ & 1.19 $\pm$ 0.04 & 18.9 & 0.51 $\pm$ 0.01 \\
buffer & 0.48 $\pm$ 0.01 & 9.45 & 0.44 $\pm$ 0.03 \\
lipid headgroup$_{inner}$ & 0.67 $\pm$ 0.02 & 12.9 $\pm$ 0.3 & 0.44 $\pm$ 0.03  \\
lipid alkylchain & 2.21 $\pm$ 0.04 & 7.1 $\pm$ 0.02 & 0.59 $\pm$ 0.03 \\
lipid headgroup$_{outer}$ & 0.88 $\pm$ 0.07 & 12.3 $\pm$ 0.4 & 0.57 $\pm$ 0.06\\
neutravidin +  & 14.2 $\pm$ 1.2 & 9.9 $\pm$ 0.1 & 1.20 $\pm$ 0.11\\
PAA-Cys5-biotin +& & & \\
\hline \hline
\end{tabular}
\end{table}

\subsection{Switching of vesicle adhesion}

\begin{figure}[!tb]
\centering
  \includegraphics[width=0.5\textwidth]{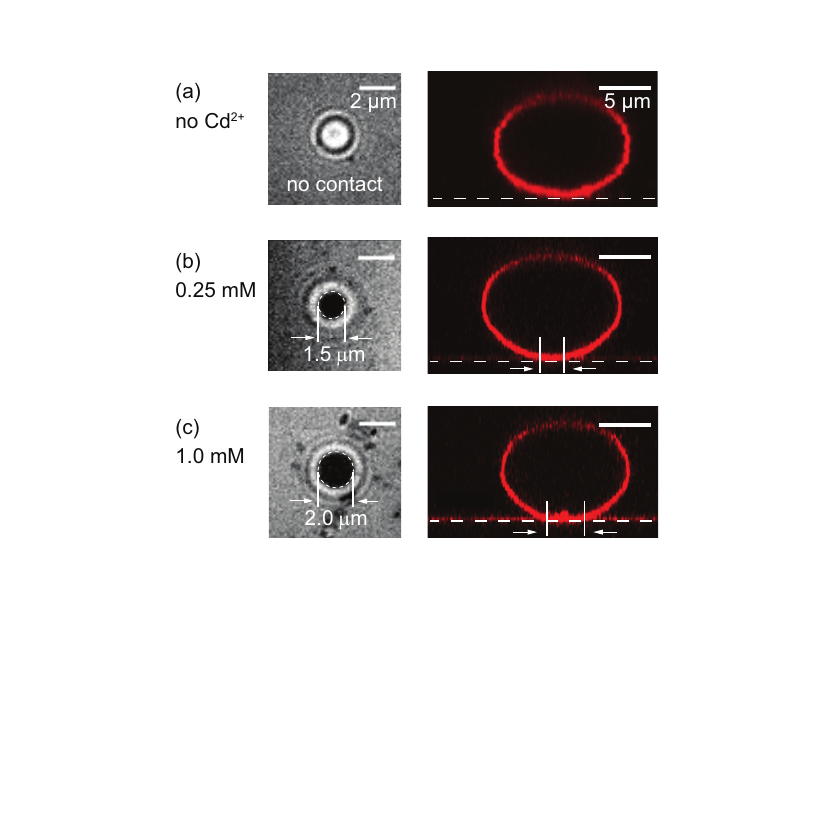}
  \caption{The confocal fluorescence images (side views) and RICM images (bottom views) of DOPC vesicles on PAA-Sys5 brushes at [Cd$^{2+}$] = 0 mM, 0.25mM and 1.0 mM, respectively.}
  \label{fgr:sideview}
\end{figure}
\autoref{fgr:sideview} shows the confocal fluorescence images (side views) and RICM images (bottom views) of DOPC vesicles on PAA-Cys5-biotin brushes at [Cd$^{2+}$] = 0 mM and 1.0 mM, respectively. Here, the global shape of the vesicle was reconstructed from confocal stacks, and the cross-sectional side view was extracted by slicing the vesicle in the middle plane. In \autoref{fgr:sideview}a (top), the confocal image taken in the absence of Cd$^{2+}$ indicates that the vesicle is near the surface by the density difference in liquid inside and outside the vesicle ($\Delta \rho = - 40$ kg/m$^3$), but no clear contact to the surface (broken line) can be identified. To verify whether the vesicle adheres on the brush surface, we observed the vesicle by RICM under the same condition (\autoref{fgr:sideview}a, left panel). The bright and fluctuating signals near the vesicle center imply that the vesicle does not adhere, and the intensity fluctuation reflects the height fluctuation \cite{albersdorfer1997adhesion,Purrucker2007Feb}. In contrast, the confocal side view taken at [Cd$^{2+}$] = 1.0 mM suggests the establishment of a “flat” contact, whose contact edges are indicated by white lines (\autoref{fgr:sideview}c, left panel). In fact, the RICM image taken under the same condition shows a stable, dark disc near the center with a diameter of about 2.0 $\mu$m. Although the confocal images and RICM images were collected on the different microscopes, all the vesicles  exhibited much less intensity fluctuation near the center at [Cd$^{2+}$] = 1.0 mM. By screening [Cd$^{2+}$] systematically, we found the onset of adhesion at even at a lower concentration, [Cd$^{2+}$] = 0.25 mM (\autoref{fgr:sideview}b Previously, Nardi \textit{et al.}~showed the change in vesicle-substrate interactions by using vesicles incorporating cationic lipids interacting with supported membranes doped with negatively charged lipids. The change in pH caused changes in surface charge density, where they observed the breakdown of Young-Dupr{\'e} type wetting by the formation of three-dimensional protrusion (blisters) \cite{nardi1998adhesion}. In contrast, our polymer brushes are able to switch lipid vesicles from non-wetting to partial wetting state with a subtle stimulus that does not not significantly alter pH or electrostatic screening. Moreover, compared to other stimulus responsive brushes such as thermoresponsive poly(N-isopropylacrylamide) (PNIPAAm) \cite{NISTOR2014114}, our brushes can switch the conformation without changing temperature, which helps us avoid hydrodynamic perturbation by thermal convection.

\subsection{Experimental determination of vesicle-brush interactions}
\begin{figure}[!tb]
\centering
  \includegraphics[width=0.5\textwidth]{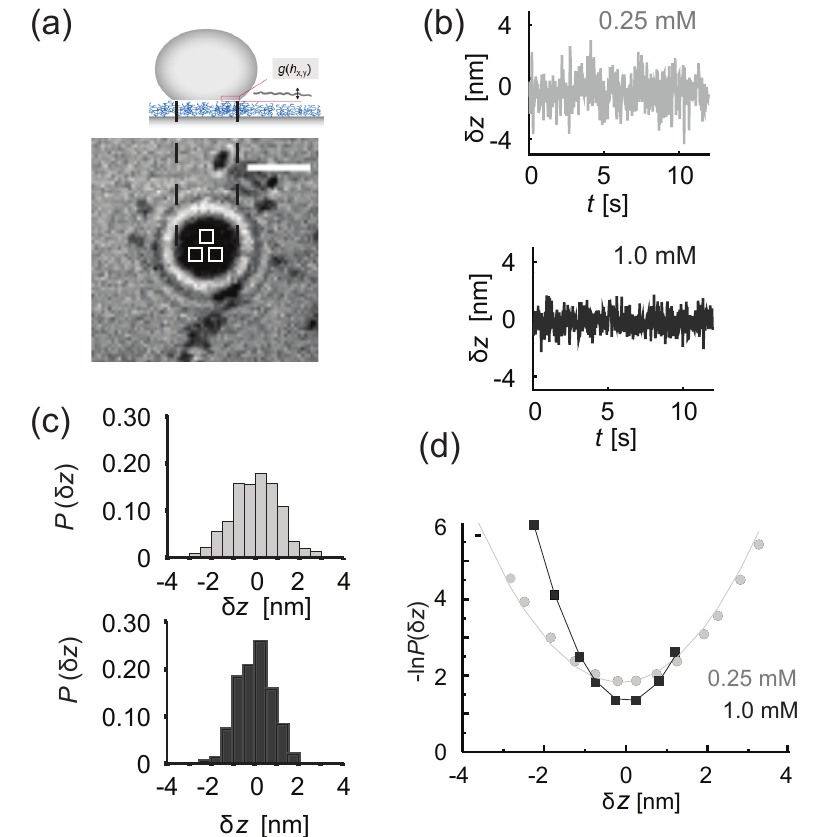}
  \caption{Membrane-substrate interaction, $V(z)$, between vesicles and brushes. (a) Schematic illustration of a vesicle adhered to polymer brushes (top) and microinterferometry image highlighting the contact zone in black (bottom). The intensity fluctuation in three independent locations (3 $\times$  3 pixels  each, indicated by white boxes) were converted to the height fluctuation following \autoref{eqn:RICM_heigth} Scale bar: 2 $\mu$m. (b) The fluctuation of membrane height $\delta z$ was monitored over time at [Cd$^{2+}$] = 0.25 mM (grey) and 1 mM (black). (c) The probability function of fluctuation amplitude of the particle $P(\delta z)$ calculated from the data presented in panel d). Under thermodynamic equilibrium, $-\ln P(\delta z)$ can be well approximated as a parabolic function around its minimum, yielding the potential curvature, $V''$, corresponding to an effective ``spring constant''.}
  \label{fig_interfacial_potential_exp}
\end{figure}

To quantitatively determine the effective membrane-substrate potential between vesicles and brushes, we analyzed the RICM movies captured at 30 ms per frame. To avoid the smear of height fluctuation due to the macroscopic membrane undulation and the camera noise \cite{bruinsma1995physics,Schmidt2014May}, we randomly selected small regions ($3 \times 3$ pixels, 3 locations) and monitored the mean intensity fluctuation inside each region (\autoref{fig_interfacial_potential_exp}a). \autoref{fig_interfacial_potential_exp}b and \autoref{fig_interfacial_potential_exp}c show the height fluctuations monitored over time, $h(t)$, and the distribution of fluctuation amplitudes, $P(h)$, measured at [Cd$^{2+}$] = 0.25 mM (grey) and 1.0 mM (black), respectively.

The characteristics of the membrane-substrate potential, $V(z)$, can be obtained from the probability, $P(\delta z)$, of local distance fluctuations, $\delta z=z-\langle z \rangle$, between membrane and substrate. $P(\delta z)$ is plotted in \autoref{fig_interfacial_potential_exp}d and is well approximated by a Gaussian distribution \cite{derjaguin1987surface, swain1999influence, tanaka2013physics}, with zero mean and variance $\langle h^2 \rangle$. In the following, we consider an almost planar membrane patch, $|\nabla z| \ll 1$, in contact with the substrate and use the Monge representation
\begin{equation}
{\cal H}_0[z] = \int \rd x \rd y\; \left\{V(z) + \Delta \rho g z + \gamma + \frac{\gamma}{2} (\nabla z)^2 + \frac{\kappa}{2} (\triangle z)^2 \right\} 
\label{eqn:Monge}
\end{equation}
Quadratically expanding the membrane-substrate interaction around its minimum,
\begin{equation}
V(z) = -\Delta \gamma_w + \frac {1}{2} V''\delta z^2  + {\cal O}(\delta z^3)
\label{eqn:derjaguin}
\end{equation}
with $V''= \left.\frac{\partial^2V} {\partial z^2}\right|_{\delta z=0}$ and assuming that the membrane tension is negligible, $\gamma\approx0$, we obtain
\begin{equation}
\frac{{\cal H}_0[\tilde z]}{L^2} = -\Delta\gamma_w +  \Delta \rho g \tilde z_{\bf 0} + \frac{1}{2}V''\tilde z_{\bf 0}^2+ \frac{1}{2}\sum_{{\bf q}\neq0} \left( V'' + \kappa q^4 \right) |\tilde z_{\bf q}|^2 
\end{equation}
where $\tilde z_{\bf q}$ denotes the two-dimensional Fourier transform of $z(x,y)$. On short length scales, the bending rigidity, $\kappa$, dominates the fluctuation spectrum, whereas on long scales the membrane-substrate potential, $V$, dictates the fluctuation behavior of the bound membrane. The crossover length scale is given by the parallel correlation length, $\xi_\|=\sqrt[4]{64\kappa/V''}$. 

The local distance, $\delta z=z-\langle z \rangle = z-\tilde z_{\bf 0}$, is given by $\delta z=\sum_{{\bf q}\neq 0}\tilde z_{\bf q}$, and its variance takes the form \cite{Helfrich84,Seifert1990Oct}
\begin{equation}
\langle \delta z^2 \rangle  = \sum_{{\bf q}\neq0} \langle |\tilde z_{\bf q}|^2\rangle = \sum_{{\bf q}\neq0} \frac{k_{\rm B}T}{L^2\left(V''+\kappa q^4 \right)}    = \frac{k_{\rm B}T}{8\sqrt{V''\kappa}} \label{eqn:zfluc}
\end{equation}
Using the definition of the parallel correlation length, $\xi_\|$, we can rewrite\autoref{eqn:zfluc} in the form
\begin{equation}
-k_{\rm B}T \ln P(\delta z) = \frac{1}{2} V'' \xi_\|^2 \delta z^2 + \mbox{const}  
\end{equation}
\ie, the height fluctuations of a point in the membrane are like those of an uncorrelated particle with a particle-substrate contact area, $\xi_\|^2$.

This analysis allows us to estimate the curvature, $V''$, of the membrane-substrate interactions at its minimum from the local height fluctuations of an adhered membrane patch. To this end, we fitted $-\ln P(\delta z)$ in \autoref{fig_interfacial_potential_exp} by a parabola and determined $\langle \delta z^2 \rangle=1.458$ nm$^2$ and $0.555$ nm$^2$ for Cd$^{2+}$ concentration of $0.25$~mM and $1.0$~mM, respectively.The bending rigidity of vesicles, $\kappa$ = 24.3 $k_BT$, was taken from the previous work \cite{matsuzaki2017adsorption}. Thus the curvature of the membrane-substrate interaction takes the values $V''=(k_{\rm B}T)^2/[64 \kappa \langle \delta z^2\rangle^2]=3.02\cdot 10^{-4}k_{\rm B}T/$nm$^4$ and $2.09 \cdot 10^{-3}k_{\rm B}T/$nm$^4$

As shown in Figure 4c, $\delta z$, of thermal fluctuations, $P(\delta z)$, is well approximated by a Gaussian and thus the shape of potential in Figure 4d resembles the red curve presented in \autoref{fig:1}. In contrast, using vesicles adhering on the micropatterned steps, Schmidt \textit{et al.}~showed that the membrane-substrate potentials of vesicles becomes nonharmonic, \cite{Schmidt2014May} which takes a similar shape as the blue curve presented in \autoref{fig:2}. The main difference between the two experimental systems is that the amplitude of height fluctuation on our PAA-Cys5 brushes are several nm, indicating that the membrane is sharply confined in the close vicinity of the potential minimum. On the other hand, the membrane on micropatterned steps fluctuates tens of nm where the long-range interactions can be detected due to a large membrane-substrate distance ($\sim 100$ nm). It should be noted that the microinterferometric analysis of membrane height fluctuation enables to characterize the potential curvature $V''$ corresponding to the “spring constant” of a harmonic oscillator but not the absolute potential energy minimum.

\subsection{Thermodynamics of adhesion in the absence of buoyancy, $\tilde \epsilon_B=0$}
\begin{figure}[!tb]
\centering
\includegraphics[width=0.45\textwidth]{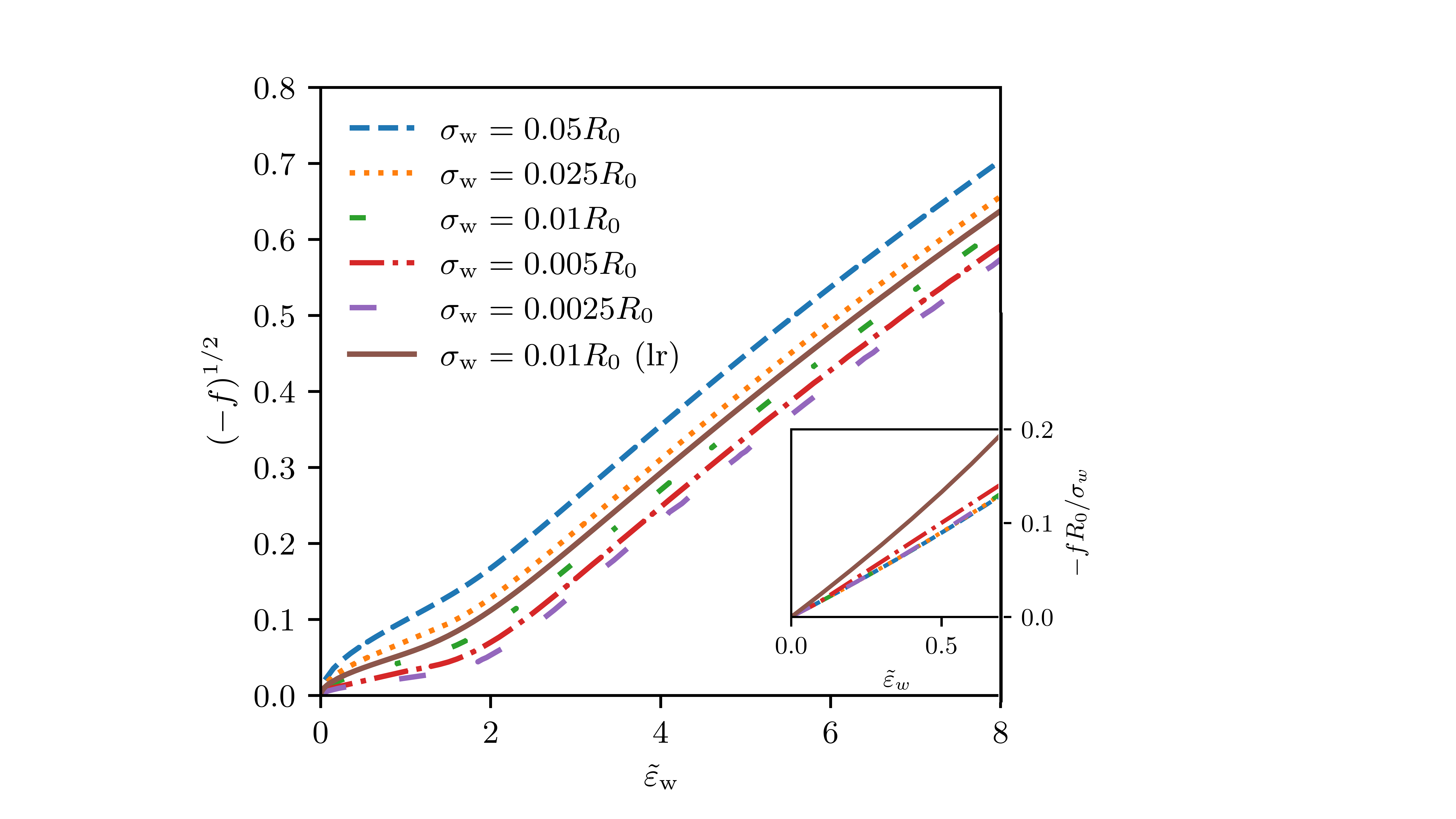}
  \caption{Root of the adsorption energy, $\sqrt{-\tilde f}$, as a function of the interaction strength, $\tilde\epsilon_w$, for the short-range and long-range  potentials, $V_c$ (dashed lines) and $V_w$(full lines), in the absence of buoyancy, $\tilde \epsilon_B=0$. For the short-range potential widths, the width, $\sigma_w$, of the membrane-substrate interaction is varied as indicated in the key. For a contact potential with $\sigma_w \to 0$, one expects $\sqrt{-\tilde f} \sim (\tilde\epsilon_w-2)$ at the second-order transition. inset: Dependence of the adsorption energy, $f$, on $\tilde \epsilon_w$ and $\sigma_w/R_0$, as expected in the pinned state, see \autoref{eqn:pin}.}
  \label{fig:2}
\end{figure}

In \autoref{fig:2} we present the dependence of the reduced adsorption energy, $\tilde f$, as a function of the adhesion strength, $\tilde \epsilon_w$, without buoyancy, $\tilde \epsilon_B=0$. For fixed $\tilde\epsilon_w$ and type of potential, the attraction increases with $\sigma_w$,  \ie $-\tilde f$ increases with $\sigma_w$. The membrane-substrate interaction is qualitatively similar for short-range and long-range potentials. The adhesion energy $-\tilde f$ of the long-range potential with width $\sigma_w=0.01R_0$ is slightly higher than of the short-range potentials with the same width and remains smaller than that of a short-range potential with the increased width, $\sigma_w=0.025R_0$. Importantly, the adsorption energy, $-\tilde f$, remains positive for all $\tilde \epsilon_w>0$ for membrane-substrate interactions of finite widths, $\sigma_w>0$. 

For a contact potential, $\sigma_w\to0$, we expect a second-order adsorption transition with \cite{Seifert1990Oct, Lipowsky1991Sep, Tordeux02b}
\begin{equation}
-\tilde f \sim (\tilde \epsilon_w-\tilde \epsilon_{wc})^2\qquad \mbox{with} \quad \tilde\epsilon_{wc} = 2 \label{eqn:adsc}
\end{equation}
Indeed, in the interval $3 \lesssim \tilde \epsilon_w \lesssim 8$ the behavior is compatible with a linear dependence of $\sqrt{-\tilde f}$ on $\tilde \epsilon_w$. The energy of an unbound vesicle, $\tilde f=0$, however, is not approached at $\tilde \epsilon_w=2$. Instead, the vesicle adopts a "pinned state"  \cite{Seifert1991Jun, Kraus1995Nov} for $\tilde \epsilon_w \to 0$, where the vesicle remains almost spherical and touches the substrate at a point. In this state, the vesicle can benefit from the finite-range attraction between membrane and surface even without deformation.

In this pinned-vesicle regime, $0 \leq \tilde \epsilon_w < 2$, we use Derjaguin's approximation to obtain an upper bound of the vesicle's energy as the energy of a spherical vesicle interacting with a planar substrate \cite{Derjaguin1934Nov} 
\begin{equation}
- \frac{\rd }{\rd z} 8\pi \kappa\tilde f = 2\pi R_0 V(z) - \frac{4\pi}{3}R_0^3  \Delta \rho g  
\end{equation}
where $z=z_{\rm cm}-R_0/2$ is the (closest) distance between the vesicle's membrane and the substrate. The minimal energy is obtained at height, $z_{\rm min}$, determined by the condition, $V(z_{\rm min})=-2\kappa\tilde\epsilon_B/(3R_0^2)$. In the absence of buoyancy, we obtain $z_{\rm min}=-\sigma_z/2$ for the finite-range membrane-substrate interaction, $V_c$. Integration yields the upper bound
\begin{equation}
\tilde f = \int_{z_{\rm min}}^\infty \frac{\rd z}{R_0}\; \left(\frac{V(z)R_0^2}{4\kappa}+\frac{\tilde \epsilon_B}{6}\right)  
\end{equation}
In the absence of buoyancy, $\tilde \epsilon_B=0$, we obtain for the short-range potential, $V_c$, the bound
\begin{equation}
-\tilde f \gtrsim \frac{1}{6} \tilde \epsilon_w \left(\frac{\sigma_w}{R_0}\right)
\label{eqn:pin}
\end{equation}
The data collapse in the inset of \autoref{fig:2} confirms this behavior that corresponds to a first-order adhesion transition. Thus, for any finite width of the membrane-substrate interaction, $\sigma_w>0$, the adsorption transition is of first order and occurs at $\tilde \epsilon_w=0$. A critical adsorption transition at $\tilde \epsilon_w=2$ only occurs in the singular limit, $\sigma_w/R_0 \to 0$.

The adhesion transition can also be observed by monitoring the contact area of the adsorbed vesicle. There is no singularity at the edge of the contact zone but the vesicle shape gradually detaches from the substrate (\textit{vide infra}). We can define a thermodynamic contact area \textit{via} the first derivative of the adhesion energy, $H={\cal H}[\psi^*,z_0^*,L_s^*]$, with respect to the adhesion strength, $\Delta \gamma_w$.
\begin{eqnarray}
\frac{\partial {H}}{\partial \Delta\gamma_w} &=& \frac{{\cal H}_w}{\Delta \gamma_w} + \int_0^{L_s} \rd s\; \underbrace{\left.\frac{\Delta {\cal H}[\psi,z_0,L_s]}{\delta \psi}\right|_*}_{=0} \frac{\partial \psi^*(s)}{\partial  \Delta\gamma_w} \\
&&+ \left.\frac{\partial {\cal H}}{\partial z_0}\right|_*\frac{\partial z_0^*}{\partial \Delta \gamma_w} + \left.\frac{\partial {\cal H}}{\partial L_s}\right|_*\frac{\partial L_s^*}{\partial \Delta \gamma_w} \nonumber \\
\tilde A_w^{\rm th} \equiv \frac{A_w^{\rm th}}{R_0^2} &=&  \frac{1}{R_0^2} \frac{\partial {H}}{\partial \Delta\gamma_w} = \frac{\sfrac{{\cal H}_w}\kappa}{\tilde \epsilon_w} 
\label{eqn:Ath} 
\end{eqnarray}
Alternatively, we can geometrically identify the radius, $r_c^{\rm geo}$, of the contact zone by the location of the maximum curvature, $C_1(s)=\sfrac{\rd \psi}{\rd s}$
\begin{equation}
r_c^{\rm geo} = r\left(\mbox {arg max}_s \sfrac{\rd \psi}{\rd s}\right) \label{eqn:rcgeo}
\end{equation}
and obtain a geometric area of the contact zone
\begin{eqnarray}
\tilde A_w^{\rm geo} &=& \frac{\pi (r_c^{\rm geo})^2}{R_0^2} \label{eqn:Ageo}
\end{eqnarray}

\begin{figure}[tbh]
\centering
  \includegraphics[width=0.4\textwidth]{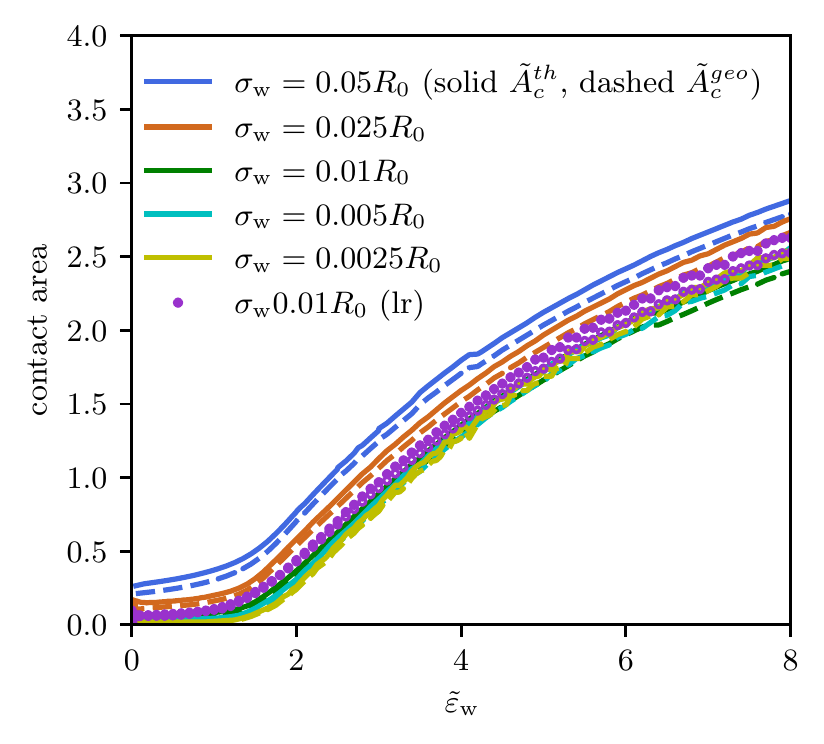}
  \caption{Thermodynamic and geometric contact area, $\tilde A_c^{\rm th}$, and $\tilde A_c^{\rm geo}$, as a function of adhesion strength, $\tilde \epsilon_w$, for short-range and long-range potentials of varying width, $\sigma_w$.}
  \label{fig:3}
\end{figure}

These estimates of the contact area, according to \autoref{eqn:Ath} (solid lines, open symbols) and \autoref{eqn:Ageo} (dashed lines, filled symbols) are shown in \autoref{fig:3}. Both definitions exhibit qualitatively similar behaviors but $\tilde A_c^{\rm geo}$ is slightly but consistently larger than $\tilde A_c^{\rm th}$. The so-defined contact area continuously increases with $\tilde \epsilon_w$, and it approaches a constant value for $\tilde \epsilon_w \to 0$ in the pinned state. For a repulsive surface, $\tilde \epsilon_w<0$, the vesicle is unbound and the contact area vanishes. The discontinuity of the contact area at $\tilde \epsilon_w=0$ marks the first-order adhesion transition. 

For small ratios $\sigma_w/R_0$ between the width of the attractive membrane-substrate potential and the vesicle's radius, the dependence of the contact area on $\tilde \epsilon_w$ changes from being independent from the adhesion strength to increasing with $\tilde \epsilon_w$ around $\tilde \epsilon_{wc}=2$ -- this is the signature of the second-order adhesion transition that emerges in the limit $\sigma_w/R_0\to 0$ and results in $\tilde A \sim \tilde \epsilon_w-2$. For the ratios considered here, $\tilde \epsilon_w/R_0\geq 0.0025$, however, there still remain significant deviations from a critical adsorption transition. By the same token, we also do not observe a jump singularity in the thermodynamic response function,
\begin{eqnarray}
\tilde c_w^{\rm th} = \frac{1}{\kappa}\frac{\partial^2 {H}}{\partial \Delta\gamma_w^2} = \frac{\partial }{\partial \tilde \epsilon_w} \left( \frac{\sfrac{{\cal H}_w}{\kappa}}{\tilde \epsilon_w} \right)
\end{eqnarray}
that is the signal of the second-order transition.

\subsection{Vesicle shape in the absence of buoyancy}
\begin{figure}[!htb]
\centering
  \includegraphics[width=0.5\textwidth]{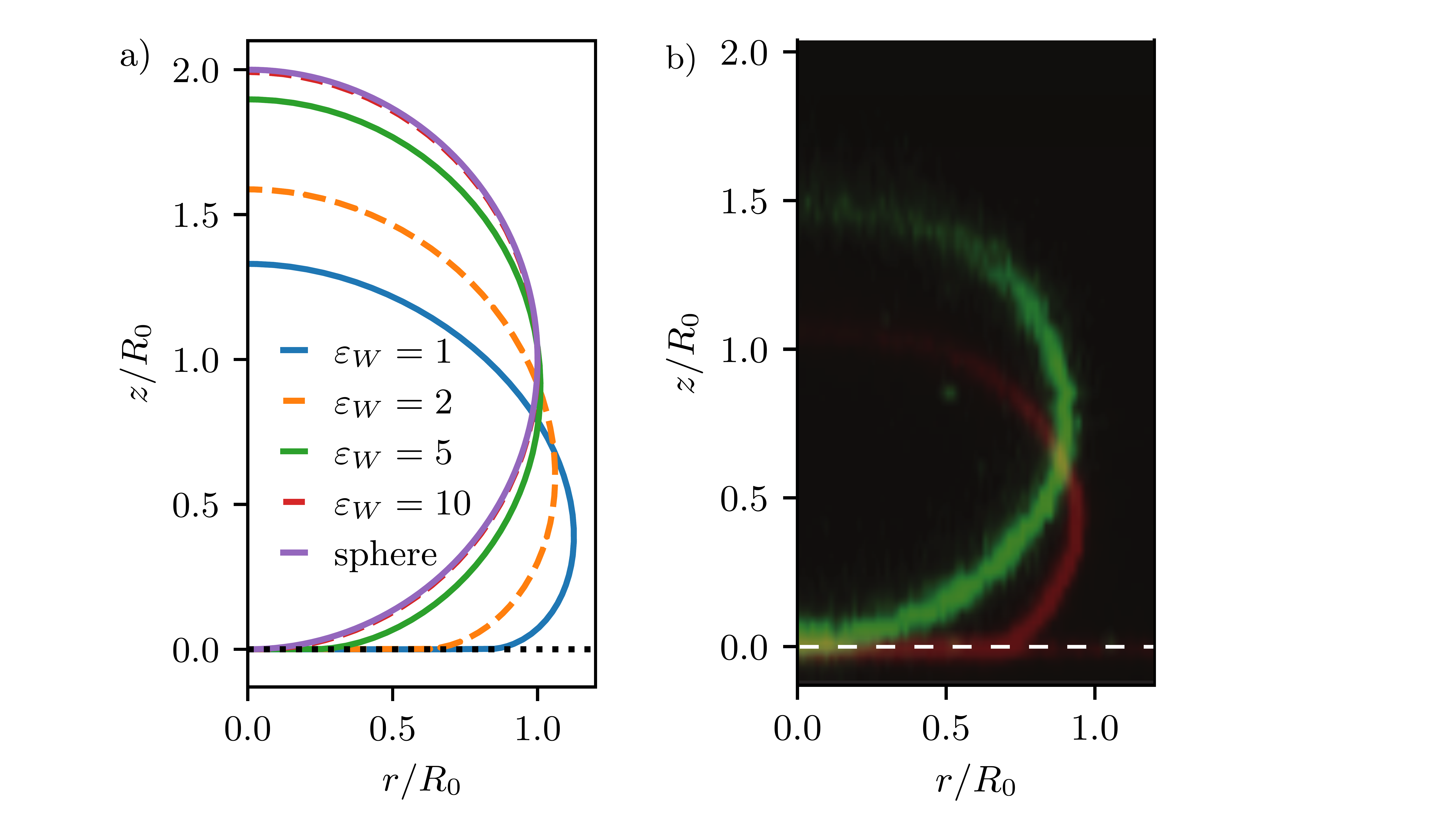}
  \caption{Vesicle shape for long-range potential, $V_w$, with scale $\sigma_w=0.002R_0$ as a function of adhesion strength, $\tilde \epsilon_w$. a) Shape of the axially symmetric vesicle, $r(s),z(s)$, for various $\tilde \epsilon_w$ as indicated in the key and comparison to a spherical vesicle. b) Vesicle shapes obtained in the absence (green) and presence (red) of attractive interactions by experiments showing the same qualitative behaviour.}
  \label{fig:comp}
\end{figure}

\begin{figure}[!htb]
\centering
  \includegraphics[width=0.5\textwidth]{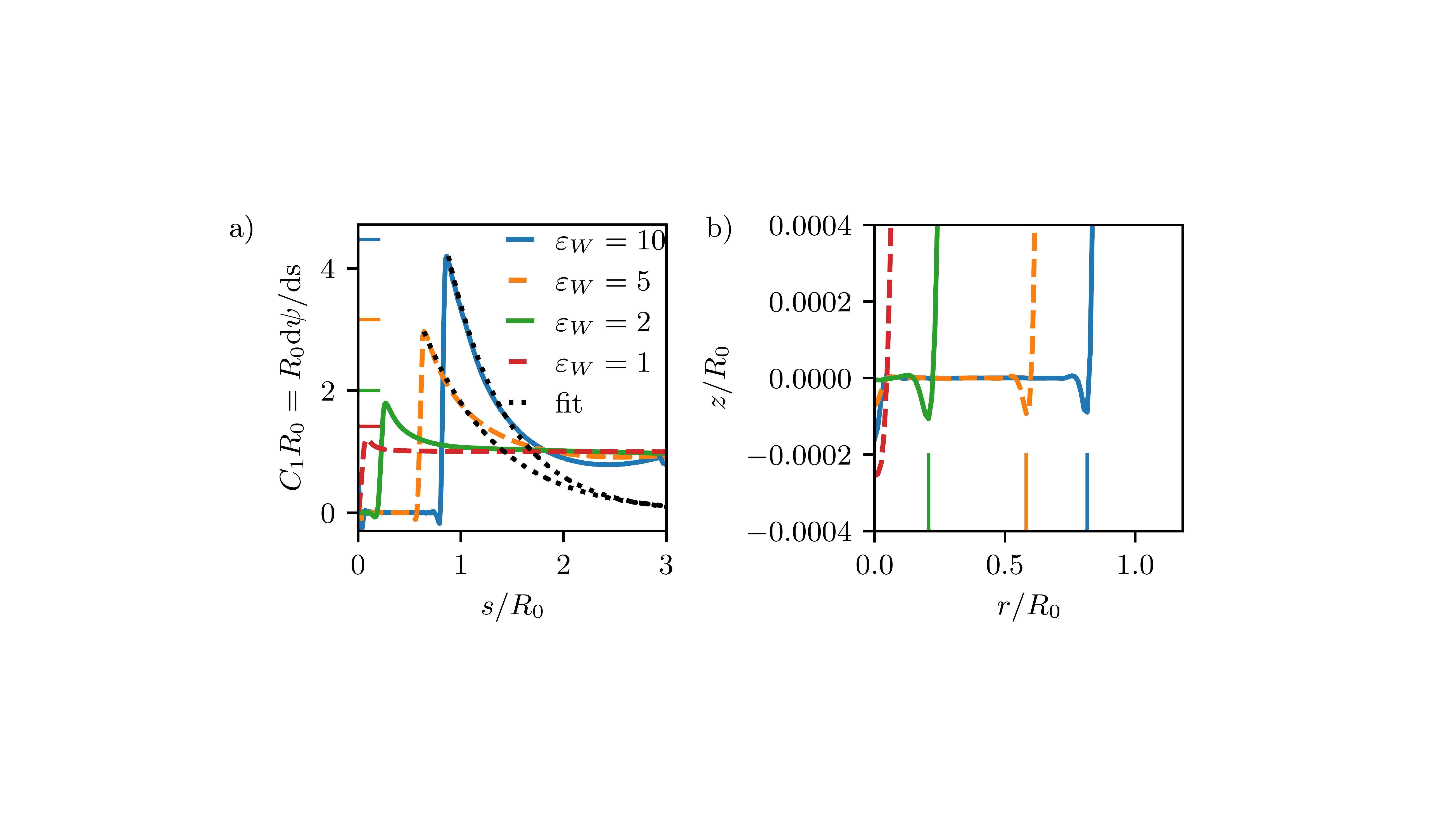}
  \caption{Vesicle shape for long-range potential, $V_w$, with scale $\sigma_w=0.002R_0$ as a function of adhesion strength, $\tilde \epsilon_w$ as indicated in the key. a) First principle curvature, $C_1(s)=\frac{\rd \psi}{\rd s}$, as a function of the arc-length parameter, $s$. Note the negative values of the curvature at the edge of the contact zone. The lines at the ordinate axis indicate the maximal-curvature estimate, $C_{1{\rm max}}R_0 = \sqrt{2\tilde \epsilon_w}$, according to the transversality condition, \autoref{eqn:transversality}. (The dotted line presents fits, $C_1(s)=C_{1{\rm max}} \exp(-[s-s_{\rm max}/\lambda_{\rm E}])$ beyond the maximum at $s_{\rm max}$, analog to \autoref{eq:bruinsma2}.) b) Detail of the vesicle shape at the edge of the contact zone, exhibiting a dent, \ie, a ring  with $z(s)<0$. The lines mark the geometric radius of the contact zone, \autoref{eqn:rcgeo}, obtained by the position of maximal curvature.}
  \label{fig:4new}
\end{figure}

\autoref{fig:comp}a) presents the vesicle shape for $V_w$ with $\sigma_w=0.002R_0$ for various adhesion strengths, $\tilde\epsilon_w$. Upon increasing $\tilde\epsilon_w$, the vesicle spreads on the substrate. For small $\tilde\epsilon_w=1$ -- the pinned state --  the vesicle shape is very close to a sphere, yet the vesicle benefits from the long-range attraction. As shown in \autoref{fig:comp}b, the confocal side views of vesicles in the absence (green) and presence (red) of attractive interactions qualitatively exhibited good agreement.

\autoref{fig:4new} b depicts the detail of the vesicle shape in the contact zone, where one can appreciate a small dent at the edge of the contact zone. Such a nonmonotonic behavior of the distance, $z(s)$, between membrane and substrate results from the simultaneous optimization of the adhesion energy and bending energy. The width of this dent increases with $\sigma_w$. Arrows indicate the geometric radius of the contact area, extracted from the maximum of the first principle curvature according to \autoref{eqn:rcgeo}. One can observe that it provides a rather faithful estimate of the edge of the contact zone.

The first principle curvature, $C_1(s)$, along the vesicle is shown in \autoref{fig:4new}a). For a contact potential, $\sigma_w \to 0$, the curvature jumps from $0$ inside the contact zone to a finite value that is dictated by \cite{Seifert1990Oct, Lipowsky1991Sep}
\begin{equation}
(C_{1{\rm max}}R_0)^2 = 2\tilde\epsilon_w \qquad \mbox{transversality condition} \label{eqn:transversality} 
\end{equation}
This boundary condition at the edge of the contact zone relates the membrane geometry, $C_1$, to the thermodynamic strength of adhesion, $\tilde \epsilon_w$. For $\sigma_w>0$, there is no jump singularity of $C_1$ but the curvature exhibits a rapid, sigmoidal variation at the edge of the contact zone. Evans suggested to use the maximal curvature, $C_{1{\rm max}}$, en lieu of the contact curvature in the transversality condition. \cite{Evans1990Jan}  Panel b of \autoref{fig:4new} reveals that the dent at the edge of the contact zone gives rise to negative $C_1$-values. Thus, the jump in $C_1$ that emerges in the limit $\sigma_w\to 0$ can alternatively be estimated by the change, $\Delta C_1 \equiv C_{1{\rm max}}-C_{1{\rm min}}$, of curvature at the edge of the contact zone. Indeed, we observe that the transversality condition is rather accurately fulfilled, even if the membrane-substrate potential has a finite range (\textit{vide infra} \autoref{fgr:trans})

\subsection{Calculation of adhesion free energy from contact curvature}
\begin{figure}[!htb]
\centering
  \includegraphics[width=0.35\textwidth]{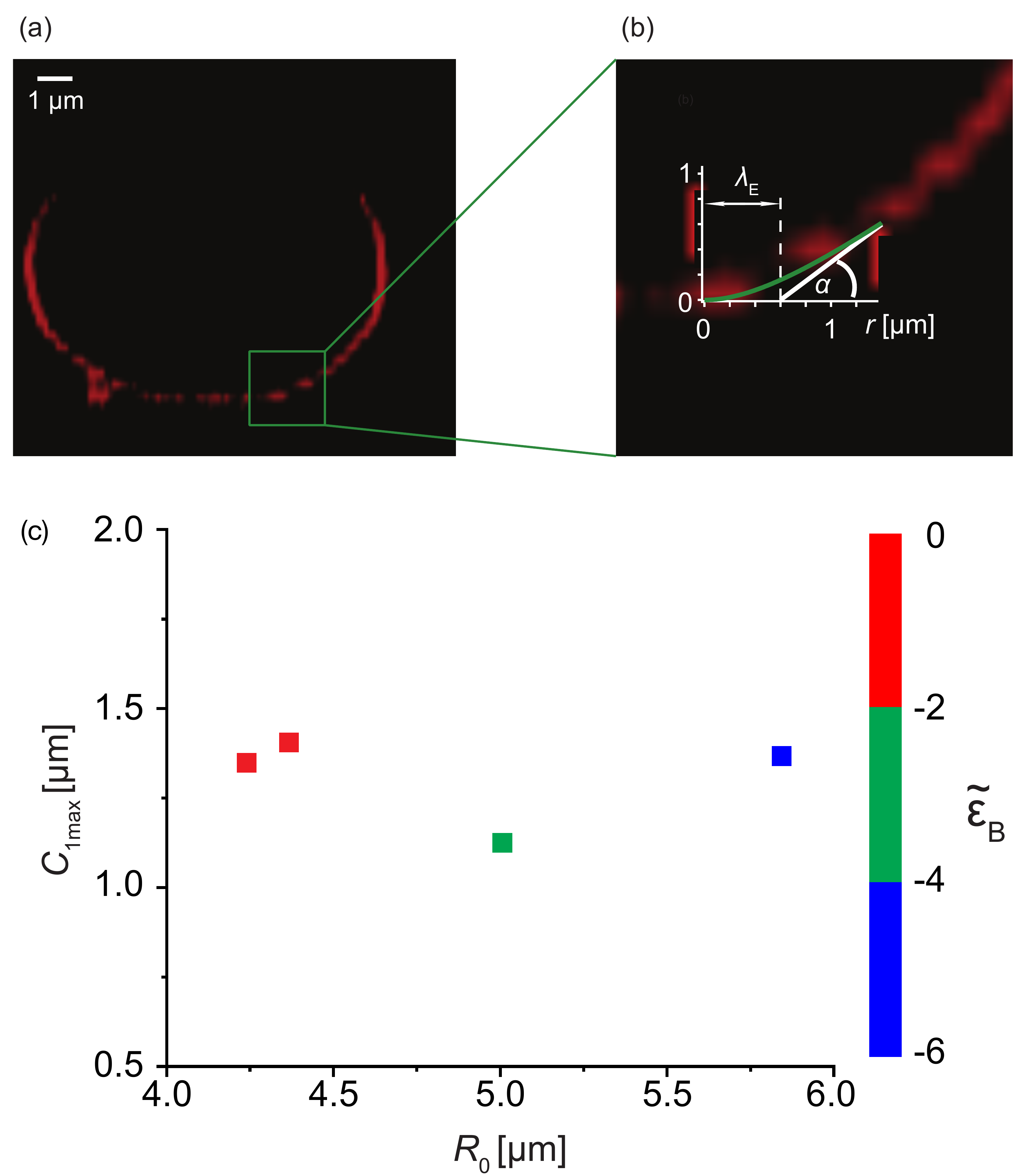}
\caption{a) A typical confocal side view of a vesicle adherent on the surface of PAA-Cys5 brushes. b) Height profile of the membrane in a higher magnification. c) Contact curvature $C_{1max}$ plotted versus $R_0$. The color code indicates the corresponding $\tilde \epsilon_B$.}
  \label{fig:felix1}
\end{figure}

\autoref{fig:felix1}a shows the shape of a vesicle near the brush substrate. The main difference between lipid vesicles and droplets of fluids is that the height profile of the membrane in the vicinity of the substrate is dominated by the bending elasticity. Only on larger scales does the geometry-dependent membrane tension, $\gamma$, become important. The crossover from bending- to tension-dominated behavior occurs on the scale of the capillary length, 
\begin{equation}
\lambda_{\gamma}=\sqrt{\frac{\kappa}{\gamma}}
\label{eqn:caplength}
\end{equation}

Outside the range of the membrane-substrate interaction, the minimization of the vesicle energy in Monge representation, \autoref{eqn:Monge}, results in a $4^{\rm th}$-order differential equation \cite{Guttenberg2000Nov}
\begin{equation}
\lambda_{\gamma}^2 \nabla^4 z - \nabla^2 z =0    
\label{eqn:elastic}
\end{equation}
\ie, $\lambda_{\gamma}$ sets the scale of the profile in the vicinity of substrate. 
For $r_{\rm E}/\lambda_{\gamma}\gg 1$, Bruinsma wrote down a one-dimensional solution\cite{bruinsma1995physics} that fulfills the boundary conditions, $z=0$, $\rd z/\rd r=0$, and $\rd^2 z/\rd r^2=\alpha/\lambda_{\gamma_{\rm E}}$ at the edge of the adhesion zone, $r=r_{\rm E}$.  Note that the fourth solution, $e^{r/\lambda}$, is not used.
\footnote{The corresponding solution with axial symmetry
\[
\frac{z(r)}{\lambda_{\rm E}} = C_{1{\rm max}} \lambda_{\rm E} \left\{
\frac{K_0(r/\lambda_{\rm E})}{K_0(r_{\rm E}/\lambda_{\rm E})}-1 + 
\frac{K_1(r_{\rm E}/\lambda_{\rm E})}{K_0(r_{\rm E}/\lambda_{\rm E})} \frac{r_{\rm E}}{\lambda_{\rm E}}\ln \frac{r}{r_{\rm E}}
\right\}
\]
has been given in Ref.~\cite{Guttenberg2000Nov}, where $K_n$ denotes the $n^{\rm th}$ modified Bessel functions of of second kind. The modified Bessel function of first kind, $I_0(r/\lambda_{\rm E})$, can also be used to construct a solution.
}
\begin{equation}
z(r) = \alpha (r-r_{\rm E} -\lambda_{\rm E}) + \alpha \lambda_{\rm E} e^{-(r-r_{\rm E})/\lambda_{\rm E}}
\label{eq:bruinsma}
\end{equation}

This expression has been employed to analyze experimental data\cite{Bruinsma2001Aug, Purrucker2007Feb}, yielding the slopes, $\alpha$, and the length scale, $\lambda_{\rm E}$. The parameter, $\lambda_{\rm E}$, is expected to coincide with the capillary length, $\lambda_\gamma$, within the scope of the approximations, $|\nabla z| \ll 1$ and $r_{\rm E}/\lambda_\gamma \gg 1$, \ie, when the fit is restricted to the ultimate vicinity of the edge of the adhesion zone and the adhesion is sufficiently strong, respectively.

The slope, $\alpha$, of the fit and the contact curvature at the adhesion edge are related \textit{via}\cite{Bruinsma2001Aug,Tordeux02b} 
\begin{equation}
C_{1{\rm max}} = \frac{\alpha}{\lambda_{\rm E}}    
\end{equation}

\autoref{fig:felix1}c shows the plot of contact curvature $C_{1{\rm max}}$ plotted vs $R_0$, taken at different  $\tilde \epsilon_B$. As shown in the figure, $C_{1{\rm max}}$ values remain almost independent from $R_0$.  

The contact curvature enables one to calculate the free energy of adhesion, using the transversality condition \autoref{eqn:transversality}:\cite{Seifert1990Oct, Lipowsky1991Sep}
\begin{equation}
\Delta \gamma_w = \frac{\kappa}{2} \left( \frac{\alpha}{\lambda_{\rm E}}    \right)^2    
\end{equation}

\begin{figure}[!htb]
\centering
  \includegraphics[width=0.4\textwidth]{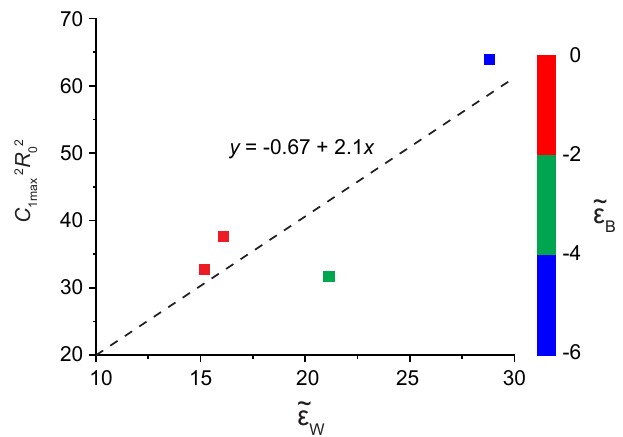}
  \caption{The first principle curvatures from experiment are plotted vs $\tilde \epsilon_w$, with indicated $\tilde \epsilon_B$. (Compare \autoref{fgr:trans}.)}
  \label{fig:felix2}
\end{figure}
In \autoref{fig:felix2}, $C_{1{\rm max}}^2R_0^2$ is plotted versus $\tilde \epsilon_w$, calculated from the average of the vesicles, whose buoyancy values are between $\tilde \epsilon_B$ = -2 and 0, and the average of corresponding $\Delta \gamma_w$ values, $\Delta \gamma_w$ = $20.3$ $k_B T/\mu m^2$. The linear fit yields the slope of 2.1 with a very small intersection (-0.67), suggesting that the deviation from the transversality condition at small $\tilde \epsilon_B$ is minor.

\begin{figure}[!tb]
\centering
  \includegraphics[width=0.45\textwidth]{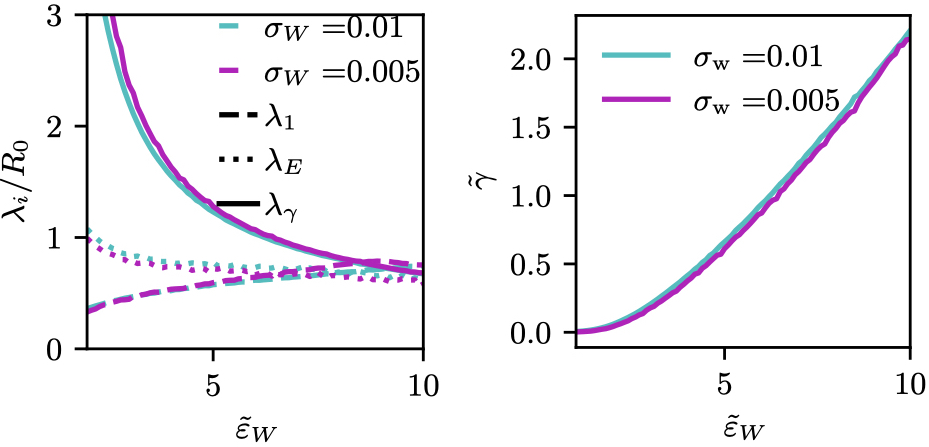}
  \caption{a) Dependence of different estimates of capillary length -- $\lambda_{\gamma}$ according to \autoref{eqn:caplength}, the extrapolation length $\lambda_1$ from a cap-shape approximation (see \autoref{fgr:model}b), and the length $\lambda_E$ extracted from the height profile in the ultimate vicinity of the edge of the adhesion zone -- on the adhesion strength, $\tilde \epsilon_w$. The figure presents data for two different interaction ranges, $\sigma_w$, as indicated in the key.  Membrane tension, $\tilde \gamma = \gamma R_0^2/\kappa$, as a function of adhesion strength, $\tilde \epsilon_w$.}
  \label{fgr:lambdaEB0}
\end{figure}

Within the Monge representation, $s \approx r$, and \autoref{eq:bruinsma} yields for the curvature, 
\begin{equation}
C_1(r) \approx \frac{\rd^2z}{\rd r^2} = \frac{\alpha}{\lambda_{\rm E}} \exp\left(-\frac{r-r_{\rm E}}{\lambda_{\rm E}}\right)   
\label{eq:bruinsma2}
\end{equation} 
for $r>r_{\rm E}$. Indeed, we qualitatively observe in \autoref{fig:4new}b that the curvature, $C_1(s)$, decays the faster (\ie, smaller $\lambda_{\gamma}$ and larger $\gamma$, see \autoref{fgr:lambdaEB0}b) the stronger the adhesion, $\tilde \epsilon_w$, is. For larger $s$, however, the Monge representation becomes inappropriate and $C_1(s)$ does not decay to zero but to a constant value that characterizes the cap-shaped, upper half of the vesicle. 

In \autoref{fgr:lambdaEB0} we systematically investigate the two geometric estimates of the capillary length -- $\lambda_1$ as depicted in \autoref{fgr:model}b and $\lambda_{\rm E}$ obtained by fitting \autoref{eq:bruinsma2} -- and compare these data with the definition, $\lambda_\gamma$, according to \autoref{eqn:caplength}.
For $\lambda_1$, we consider the $18\%$ of the data closest to $s=L_s$ and fit a sphere (where we assumed axial symmetry for the experimental data). This spherical cap is characterized by its radius, $R_{\rm cap}$, and the height of the top, $z(L_s)$. Given the radius, $r_{\rm E}$, of the edge of the adhesion zone, $\lambda_1$, is obtained.

First we determine the value, $C_{1{\rm max}}$. Then, $\lambda_{\rm E}$ is estimated from \autoref{eq:bruinsma2} by a one-parameter fit in the vicinity of $r \gtrsim r_{\rm E}$.


The comparison of the two data sets with different ranges, $\sigma_w$, of membrane-substrate interactions reveals that this local characteristics is largely independent from $\sigma_w$, similar to the behavior of the transversality condition.

The three different estimates differ for small $\tilde \epsilon_w$ and only appear to converge to a common value for large adhesion strength. Given the involved approximation, $r_{\rm E}\ll \lambda_{\gamma}$, these deviations are expected. For small adhesion strength, $\tilde \epsilon_w$, the vesicle is nearly spherical and the geometry-dependent tension is very small. Upon increasing $\tilde \epsilon_w$, we observe a rather pronounced decrease of $\lambda_\gamma$. $\lambda_{\rm E}$ obtained from the decay of the curvature at the edge of the contact zone also decreases with the adhesion strength, $\tilde \epsilon_w$, but the dependence is significantly weaker. The estimate, $\lambda_1$, that is extracted from the overall shape of the vesicle, assuming a cap shape, even displays the opposite $\tilde \epsilon_w$-dependence at small adhesion strength.  This finding highlights the challenge of accurately estimating the membrane-vesicle interactions \textit{via} analytical but approximate descriptions of the vesicle shape.

\subsection{Role of buoyancy on vesicle shape}
\begin{figure}[!tb]
\centering
  \includegraphics[width=0.45\textwidth]{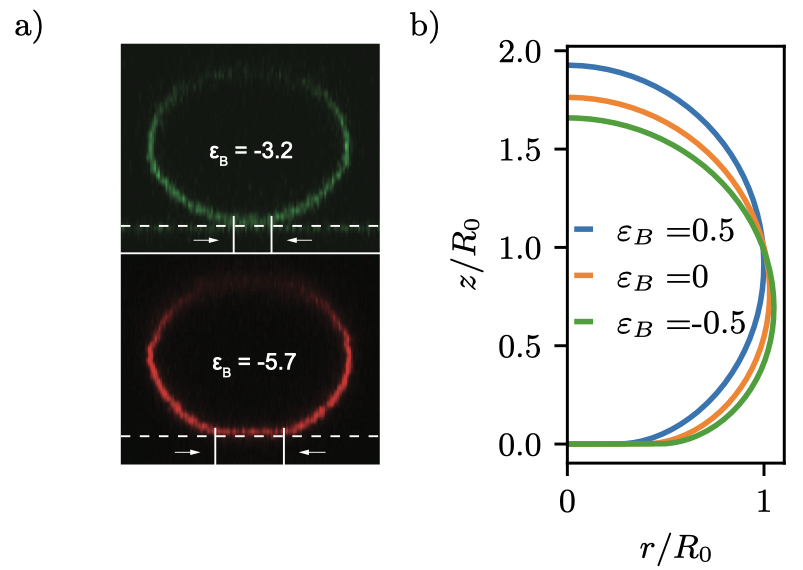}
  \caption{a) The experimentally observed vesicle shapes under two different downward buoyancies. b) Vesicle shapes as a function of buoyancy, $\tilde \epsilon_B$ at $\tilde \epsilon_w=2.5$ and $\sigma_w=0.01R_0$.}
  \label{fgr:ShapesBuoyancs}
\end{figure}
In this section, we focus on the role of buoyancy. The buoyancy significantly affects the shape of the vesicle.\cite{Kraus1995Nov} This is of importance as, in experimental set-ups, buoyant forces are regularly used to let the vesicles adhere onto the substrate. Usually, the downward buoyancy is used to push down vesicles onto the surface. To have vesicles to adhere at upward buoyancy, we flip the substrate upside-down then mount the sample of the microscope. \autoref{fgr:ShapesBuoyancs}a shows the shape of vesicles captured under two different downward buoyancy conditions,  $\tilde \epsilon_B= -3.2$ (top) and  $\tilde \epsilon_B= -5.7$ (bottom), implying that the vesicle at $\tilde \epsilon_B= -5.7$ had a larger vesicle-brush contact than at  $\tilde \epsilon_B= -3.2$. To illustrate this buoyancy effect, we compare those findings with simulated vesicle shapes (\autoref{fgr:ShapesBuoyancs}b)at a fixed $\tilde \epsilon_w=2.5$ and $\sigma_w=0.01\,R_0$ as a function of buoyancy, $\tilde \epsilon_B$. As expected intuitively, a downward buoyancy, $\tilde \epsilon_B<0$ increases the radius, $r_{\rm E}$, of the edge of the adhesion zone and reduces the height, $z(r=0)$, of the top of the vesicle. A heavy vesicle flattens nonuniformly, \ie downward buoyancy gives rise to a nonuniform rescaling of the height coordinate, $z(r)$, compared to $\tilde \epsilon_B=0$. 
Upward buoyancy, $\tilde \epsilon_B>0$, has the opposite effect, qualitatively.

Since $\tilde \epsilon_B \sim R_0^4$, the balance between bending energy and potential energy in the gravitational field exhibits a pronounced dependence on the vesicle size, $R_0$. Thus even small density differences between the interior of the vesicle and the surrounding solution give rise to large values of $\tilde \epsilon_B$ for large vesicles; these effects should be considered in a quantitative analysis.

\begin{figure}[!tb]
\centering
  \includegraphics[width=0.48\textwidth]{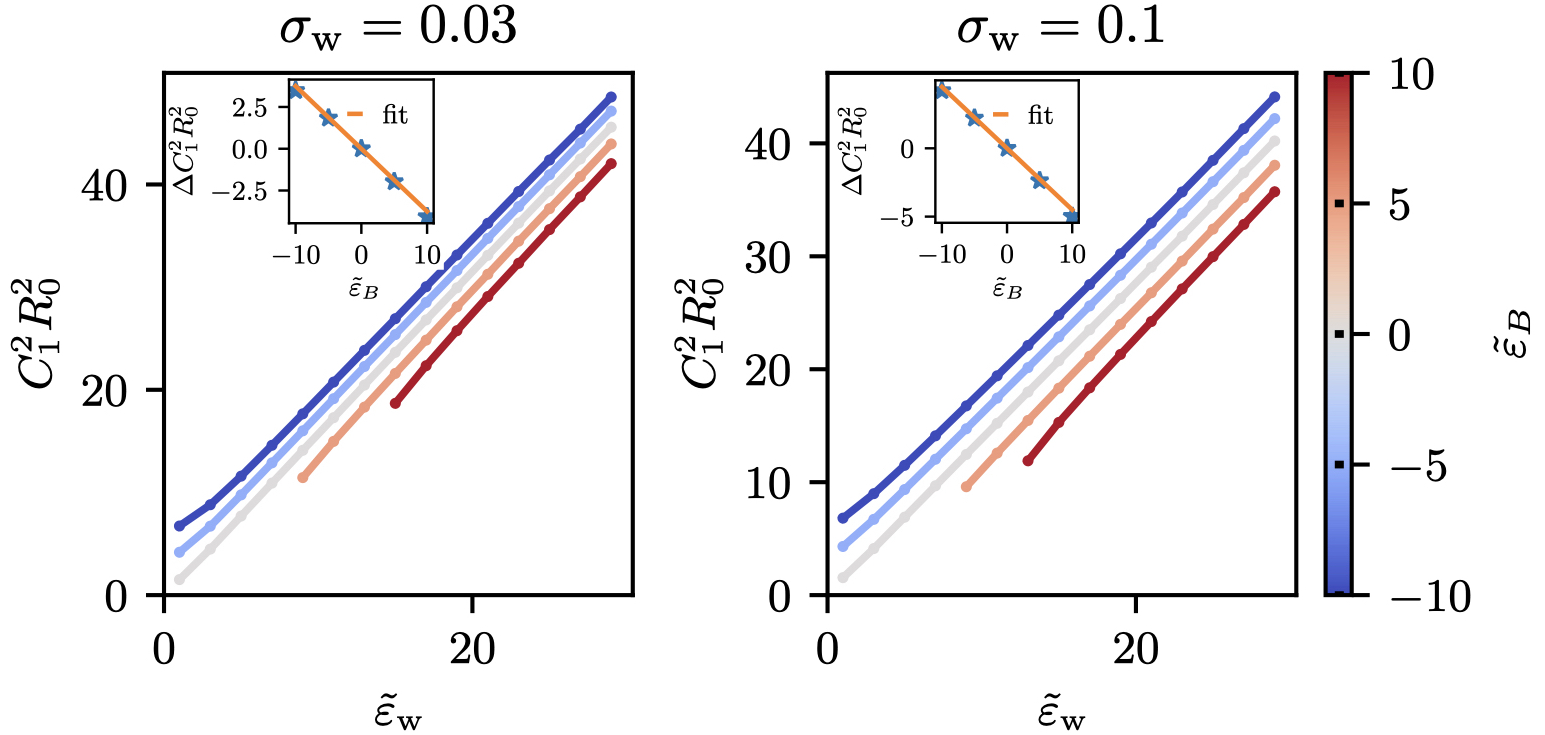}
  \caption{Transversality condition as a function of buoyancy, $\tilde \epsilon_B$. Panels a and b correspond to two distinct ranges of membrane-substrate interaction, $\sigma_w=0.03R_0$ and $0.1R_0$, respectively. The insets present the shift $\Delta (C_{1{\rm max}}R_0)^2$ as a function of $\tilde \epsilon_B$.}
  \label{fgr:trans}
\end{figure}

In the following, we focus on the shape in the ultimate vicinity of the edge of the adhesion zone. In the absence of buoyancy, we have found that the transversality condition $C_{1{\rm max}}^2 =\frac{2\Delta \gamma_w}{\kappa}$ holds for zero-ranged and finite-ranged membrane-substrate interactions. This local equilibrium balance between adhesion and curvature is a boundary condition for the elastic shape equations, such as e.g., \autoref{eqn:elastic}, and, therefore, it is independent of the vesicle size, $R_0$. 

Since we have employed the transversality condition to estimate the adhesion strength, it is important for a quantitative analysis to investigate to what extent this relation is modified by buoyancy. In \autoref{fgr:trans} we plot the square of the dimensionless maximal curvature, $C_{1{\rm max}}R_0$, as a function of the adhesion strength, $\tilde \epsilon_w$, for various values of buoyancy, $\tilde \epsilon_B$, as indicated by the color code. Even in the presence of buoyancy, the data can be well described by a linear relation 
\begin{equation}
(C_{1{\rm max}}R_0)^2 \approx \tilde \epsilon_w + \Delta (C_{1{\rm max}}R_0)^2    
\end{equation}
The slope remains unaltered but there is an offset $\Delta (C_{1{\rm max}}R_0)^2$ that depends on buoyancy. This buoyancy dependence of the off set is depicted in the insets of \autoref{fgr:trans} for two ranges of the membrane-wall interaction, $\sigma_w/R_0= 0.03$ and $0.1$, respectively. For $\tilde \epsilon_B=0$, this offset vanishes and it decreases with buoyancy. $\Delta (C_{1{\rm max}}R_0)^2$ is well parameterized by a linear dependence, \ie, $\Delta (C_{1{\rm max}}R_0)^2 \sim \tilde \epsilon_B$ for the interval of buoyancy considered. The proportionality constant depends on the functional form of the membrane-substrate interaction and we find the slopes  $-0.375$ and $-0.45$ for $\sigma_w/R_0=0.03$ and $0.1$, respectively. $\Delta (C_{1{\rm max}}R_0)^2$ effectively accounts for modification of the interplay between adhesion and bending due to gravitation. Since the transversality condition is a local balance, we hypothesize that the relevant length scale is the range of the membrane-substrate interaction. In this case, we expect that the off set, $\Delta (C_{1{\rm max}R_0})^2$, becomes negligible for zero-range interactions. This is compatible with the data presented in \autoref{fig:felix2}.

\begin{figure}[!tb]
\centering
  \includegraphics[width=0.47\textwidth]{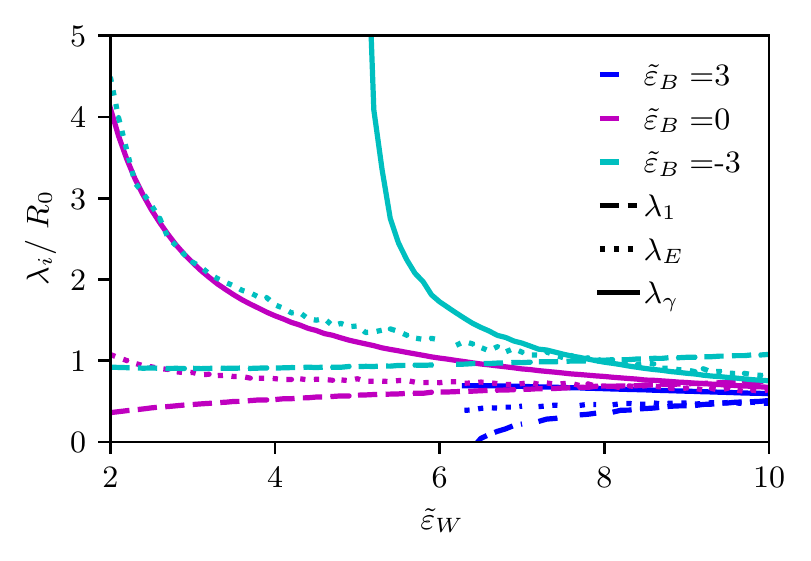}
  \caption{The extrapolation length changes significantly with the change of buoyancy as expected. However, for small downward buoyancy, the relation between $\lambda_1$ and $\lambda_2$ hardly changes. When the shape of the vesicle starts to flatten considerably due to buoyancy, the relation between the length scales approaches a straight line.}
  \label{fgr:lambda}
\end{figure}

Whereas the role of buoyancy on the transversality condition can be simply captured by an heuristic off set, $\Delta ({C_{1{\rm max}}}R_0)^2$, buoyancy has a more significant effect on the large-scale shape of a vesicle. While it has been argued that buoyancy also result in a breaking of axial symmetry \cite{Kraus1995Nov}, we limit our following considerations to axially symmetric shapes.

We start out by noting that buoyancy in the case of the wetting of liquid changes the droplet's interface from a cap shape to a pancake for large Bond number, Bo$=\Delta \rho g R_0^2/\gamma \gg 1$, with $R_0$ denoting the droplet size. In this case, a wetting angle is not well defined. Similarly, we observe in \autoref{fgr:ShapesBuoyancs} a flattening of the vesicle for downward buoyancy, $\tilde \epsilon_B>0$. Thus, the upper half of the vesicle cannot be completely described by a cap-shape because the cap radius, $R_{\rm cap}$, diverges for a flat-top vesicle. Thus, there is a significant ambiguity in extracting the geometric scale, $\lambda_1$, of \autoref{fgr:model}. 

In \autoref{fgr:lambda} we plot the two geometric scales, $\lambda_1$ and $\lambda_{\rm E}$, extracted from cap-shape fit of the top of the vesicle and obtained by fitting the decay of curvature according to \autoref{eq:bruinsma2}, respectively. We compare these with the capillary length, $\lambda_{\gamma}$ (see \autoref{eqn:caplength}). For large upward buoyancy and small adhesion strength, there are no data because the vesicles are not even metastable (\textit{vide infra \autoref{sec:phasediagram}})

The behavior is qualitatively similar to that for $\tilde \epsilon_B=0$, shown in \autoref{fgr:lambdaEB0} but, as expected, the deviations between $\lambda_1$ and $\lambda_{\rm E}$ are even larger for downward buoyancy as (i) the inappropriateness of the spherical-cap approximation and (ii) the modification of \autoref{eqn:elastic}, which is used to motivate \autoref{eq:bruinsma}, increase with buoyancy. This highlights the need to accurately model the vesicle shape in order to quantitatively estimate adhesion strength or membrane tension.

Qualitatively, the dependence of $\lambda_{\rm E}$ on buoyancy can be related to the behavior of the capillary length, $\lambda_\gamma$. For upward buoyancy, $\tilde \epsilon_B>0$, the shape of the metastable vesicle is deformed by the gravitation field even for small adhesion strengths, $\tilde \epsilon_w \lesssim 2$, in the pinned state. Thus, the membrane tension, $\gamma$, for small $\tilde \epsilon_w$ is larger than in the absence of upward buoyancy and, consequently, the capillary length, $\lambda_\gamma$, decreases with $\tilde \epsilon_B$. Since $\gamma$ does not approximately vanish for pinned vesicles, the strong increase of $\lambda_\gamma$ for $\tilde \epsilon_B \to 2$ is avoided. \autoref{fgr:lambda} shows the analog of this behavior of $\lambda_{\rm E}$. 

For downward buoyancy we observe a sign change in $\gamma$, when approaching small $\tilde \epsilon_w$ values. Below, $\lambda_{\gamma}$ is no longer defined and the gravitation field need to be included in the description, \autoref{eqn:elastic}. 

\subsection{Role of buoyancy on the adhesion transition}
\label{sec:phasediagram}
In the absence of buoyancy, $\tilde \epsilon_B=0$, there is a mean-field adhesion transition at $\tilde \epsilon_W=0$ for a finite width of the membrane-substrate interaction. In the following, we employ concepts from the description of wetting \cite{schick90, Muller2003, Indekeu1999Jun, Binder03, Bonn2009May} to analyze the thermodynamics of the system. To this end, we adopt a coarse-grained description where we characterize the system configuration by a single, scalar variable -- the center-of-mass position, $z_{\rm cm}$ of the vesicle -- instead of considering the detailed shape, $(r(s),z(s))$, of the vesicle. Thereby, we assume that for a given $z_{\rm cm}$, the shape of the vesicle  minimizes the energy, and we denote this energy by the vesicle potential
\begin{equation}
\tilde g(z_{\rm cm}) = \frac{\left.{\cal H}_0[\psi^*,z_0^*,L_s^*]\right|_{Z=z_{\rm cm}}}{8\pi\kappa}-1    
\end{equation}
where $\psi^*,z_0^*$, and $L_s^*$ minimize ${\cal H}$ under the restraint $Z=z_{\rm cm}$ (see \autoref{eqn:rzcm}). In the absence of buoyancy, $\tilde g(z_{\rm cm}) \to 0$ for $z_{\rm cm} \to \infty$. The minimum of the vesicle potential occurs at the equilibrium center-of-mass position, $z_{\rm cm}^*$, and the value of the vesicle potential at this position is the adhesion energy, $\tilde f= \tilde g(z_{\rm cm}^*)$. 

This vesicle potential, $\tilde g(z_{\rm cm})$, plays a similar role in vesicle adhesion as the interface potential, which describes the free energy of placing a unit area of liquid-vapor interface a given distance away from a solid substrate, in the wetting of liquids. This concept is very powerful for describing the surface thermodynamics of wetting and adhesion. 
In the following, an additional contribution to the vesicle potential due to Helfrich repulsion \cite{Helfrich1978Mar} or a softening of the bending rigidity due to thermal fluctuations,\cite{Peliti1985Apr} and additional finite-temperature effects \cite{Gruhn2005Jan} are ignored.

\begin{figure}[h]
\centering
  \includegraphics[width=0.45\textwidth]{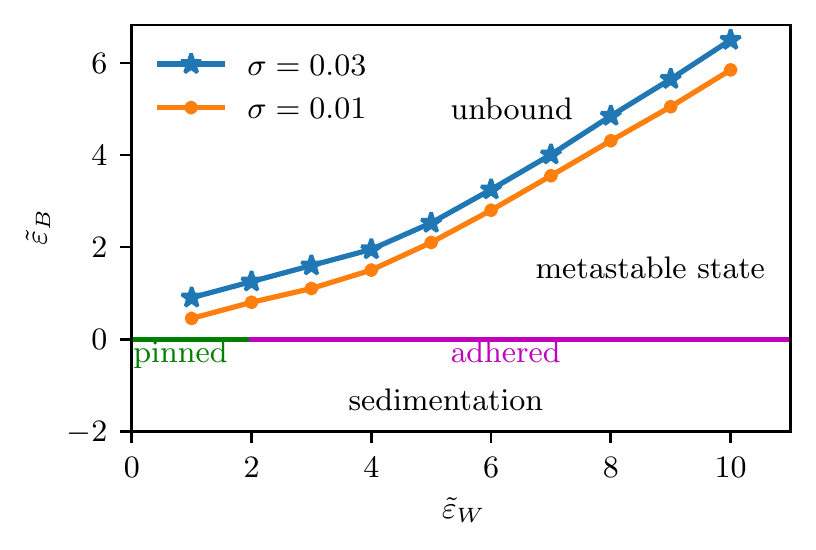}
  \caption{Mean-field adhesion diagram of vesicle adhesion as a function of adhesion strength, $\tilde \epsilon_w$, and buoyancy, $\tilde \epsilon_B$. When pinning the center of mass of the vesicle at different heights, while slowly detaching a vesicle, with upward buoyancy, an energy minimum is observed. For increasing buoyancy, this minimum will eventually disappear. Here, the first $\tilde \epsilon^{\rm spin}_B$, at which the vesicle potential does not have a local minimum, is shown for several $\tilde \epsilon_w$ for $\sigma_w/R_0=0.03$ and $0.01$. 
  } 
  \label{fgr:transition}
\end{figure}

To make a connection to the theory of wetting,\cite{schick90, Muller2003, Indekeu1999Jun, Binder03, Bonn2009May} we identify the center-of-mass position, $z_{\rm cm}$, of the vesicle with the thickness, $h$, of the wetting layer. The dewetted state, where $h$ is microscopic, corresponds to a pinned or adhered vesicle, whereas the unbound vesicle is the analog of a macroscopically thick wetting film. The vesicle potential and the interface potential are comprised of short-range and long-range contributions. The short-range contribution of the interface potential arises from the distortion of the liquid-vapor interface due to the presence of the substrate. The short-range contribution to the vesicle potential arises from (i) the bending energy of the vesicle and (ii) the short-range contribution of the direct membrane-substrate interaction, $V$. Note that the former contribution is always repulsive. In both cases -- wetting and vesicle adhesion -- the long-range contribution to the interface potential and vesicle potential, respectively, arise from van-der-Waals interactions. The corresponding power laws, however, differ (see \autoref{eqn:Vw}) because in case of wetting the entire liquid film of thickness $h$ interacts with the substrate whereas, in the case of a vesicle, only a thin membrane at a distance, $z_{\rm cm}$, interacts with the substrate. Wetting occurs only at coexistence between liquid and vapor phases. If the chemical potential differs from the coexistence value by an amount $\Delta \mu$, this difference will give rise to a linear contribution to the interface potential. Therefore we identify the role of the chemical potential difference in wetting with the buoyancy, $\tilde \epsilon_B$, in vesicle adhesion. In analogy to wetting, the vesicle adhesion transition can only occur at vanishing buoyancy. For upward buoyancy, $\tilde \epsilon_B>0$, the unbound vesicle always is the absolute minimum of the vesicle potential, whereas for downward buoyancy, $\tilde \epsilon_B<0$, the vesicle sediments onto the substrate and the vesicle potential adopts its minimum at a finite $z_{\rm cm}$ of the order $R_0$. In the case of long-range membrane-substrate interaction, \autoref{eqn:Vw}, we obtain the asymptotic behavior
\begin{equation}
\tilde g(z_{\rm cm}) \sim - \tilde \epsilon_w \left( \frac{\sigma_w}{R_0}\right)^3  \left( \frac{R_0}{z_{\rm cm}}\right)^3  - \frac{\tilde \epsilon_B}{6} \frac{z_{\rm cm}}{R_0}  \qquad \mbox{for} \quad z_{\rm cm} \gg R_0
\label{eqn:Vasym}
\end{equation}
where prefactors of order unity have been ignored. 

There is, however, one important difference between the wetting of a liquid on a substrate and the adhesion of a vesicle. Wetting phenomena refer to singularities of the excess free energy of the surface that only occur when the lateral extent of the liquid film on the substrate becomes macroscopic (thermodynamic limit). In the case of vesicle adhesion, however, the vesicle is always of finite size, $R_0$, and the limit $R_0 \to \infty$ is nontrivial as the thermodynamic control variables, adhesion strength $\tilde \epsilon_w$ and buoyancy $\tilde \epsilon_B$, depend on the vesicle size, $R_0$. Only in the mean-field approximation, where we minimize ${\cal H}[\psi,z_0,L_s]$ and ignore thermal fluctuations, can an adhesion transition occur.

From these general considerations and the asymptotic behavior, \autoref{eqn:Vasym}, we deduce the following mean-field adhesion diagram. For $\tilde \epsilon_w>0$, \ie, where the membrane-substrate interaction is attractive, the vesicle is close to the substrate in the pinned or adhered state for $\tilde \epsilon_B\leq 0$. For upward buoyancy, $\tilde \epsilon_B > 0$, the unbound state is always stable but the pinned or bound state may be metastable. This line of metastability, $\tilde \epsilon_w^{\rm spin}(\tilde \epsilon_B)$, is presented for two potential widths in \autoref{fgr:transition}. This attractive strength, $\tilde \epsilon_w^{\rm spin}$, up to which the pinned or adhered vesicle remains metastable at the substrate and can be studied by experiment, increases with the adhesion strength, $\tilde \epsilon_w$. Similar to the behavior of the contact area, we observe a gradual change between the pinned and adhered states. Also, we observe that a more extended membrane-substrate interaction (at fixed $\Delta \gamma_w \sim \tilde \epsilon_w$), stabilizes the metastable vesicle in the vicinity of the substrate.

Inside this metastable regime, there exists a finite energy barrier between the metastable pinned/adhered state and the thermodynamically stable, unbound state. This energy barrier dictates the lifetime of the  vesicle at the substrate and remains an interesting topic for further studies.

\section{Conclusions}
We synthesized a cysteine-modified polyacrylic acid polymer, PAA-Cys5, that changes the conformation in the presence of Cd$^{2+}$ ions in a concentration dependent fashion, and grafted the brushes on supported membranes at the controlled grafting distance $\langle d \rangle =  6$ nm. Specular X-ray reflectivity data indicated the compaction of PAA-Cys5 layer at [Cd$^{2+}$] = 1 mM (\autoref{figure_XRR_SLD}), which switches the phospholipid vesicles from non-adhered to adhered state. By carefully observing the vesicle-brush contact zone by microinterferometry, we determined the critical concentration inducing the switching, [Cd$^{2+}$]  = 0.25 mM (\autoref{fgr:sideview}). The analysis of height fluctuation of membranes  and the shape of vesicles near the surface enabled us to determined the curvature of the membrane-substrate interaction, $V"(z)$  (\autoref{fig_interfacial_potential_exp}), contact curvature $C_{1{\rm max}}$, and the adhesion free energy $\Delta \gamma_w$ (\autoref{fig:felix1}). 

We have studied the shape and thermodynamics of vesicles on stimulus responsive substrates. Whereas most previous studies focus on the idealized situation of zero-ranged contact interaction in the absence of buoyancy (see Ref.~\cite{Seifert1991Jun,Kraus1995Nov} for exceptions), we pay particular attention to these aspects because they are relevant to experiments. Both aspects have only a minor influence on the maximal curvature at the edge of the adhesion zone (see \autoref{fgr:trans}), slightly shifting the transversality condition that is commonly employed to relate the contact curvature to the adhesion strength. The thermodynamics and mean-field adhesion diagram, however, differs qualitatively (as compiled in \autoref{fgr:transition}). Buoyancy play the role of the deviation of the chemical potential from coexistence in wetting and, accordingly, a adhesion transition can only occur at zero buoyancy. For upward buoyancy, $\tilde \epsilon_B>0$, unbound vesicles are stable but bound vesicles may remain metastable for sufficiently large adhesion strengths. For downward buoyancy, $\tilde \epsilon_B<0$, which is common in experiments, vesicles sediment onto the substrate and for nonrepulsive substrates there is no mean-field phase transition. To achieve the vesicle adhesion at zero buoyancy is not experimentally possible, but the experimental results obtained at small buoyancy agrees well with the transversality condition (\autoref{fig:felix2}) in agreement with \autoref{fgr:trans}.  A mean-field adhesion transition can occur for $\tilde \epsilon_B=0$. For finite-range interactions (both, short-range and long-range), the transition is of first order and occurs when the membrane-substrate interaction switches from adhesive, $\tilde \epsilon_w>0$, to repulsive, $\tilde \epsilon_w<0$. A second-order adhesion transition \cite{Seifert1990Oct, Lipowsky1991Sep, Seifert1991Jun, Tordeux02b} at a finite $\tilde \epsilon_{wc}=2$ only occurs as singular limit of vanishing interaction range, $\sigma_w \to 0$. 

\section*{Author Contributions}
MM and MT designed the research, acquired funding, and supervised the project. MN synthesized the polymer. FW performed the experiment and analyzed the experimental data. LW supported the data analysis, performed simulations, and devised the adhesion diagram. FW, LW, MM, and MT discussed the data, and all the authors contributed to write the manuscript. 

\section*{Conflicts of interest}
There are no conflicts to declare.

\section*{Acknowledgements}
The authors thank the Deutsche Forschungsgemeinschaft (DFG) within the priority programme SPP2171 under grant numbers Mu1674/17-1 (to MM) and Ta253/14-1 (to MT), and the JSPS KAKANHI under the grant number JP19H05719 (to MN) for financial supports. LW and MM gratefully acknowledge the Gauss Centre for Supercomputing e.V. (www.gauss-centre.eu) for funding this research project by providing computing time through the John von Neumann Institute for Computing (NIC) on GCS Supercomputer JUWELS at the Jülich Supercomputing Centre (JSC).



\balance



\begin{mcitethebibliography}{70}
\providecommand*{\natexlab}[1]{#1}
\providecommand*{\mciteSetBstSublistMode}[1]{}
\providecommand*{\mciteSetBstMaxWidthForm}[2]{}
\providecommand*{\mciteBstWouldAddEndPuncttrue}
  {\def\EndOfBibitem{\unskip.}}
\providecommand*{\mciteBstWouldAddEndPunctfalse}
  {\let\EndOfBibitem\relax}
\providecommand*{\mciteSetBstMidEndSepPunct}[3]{}
\providecommand*{\mciteSetBstSublistLabelBeginEnd}[3]{}
\providecommand*{\EndOfBibitem}{}
\mciteSetBstSublistMode{f}
\mciteSetBstMaxWidthForm{subitem}
{(\emph{\alph{mcitesubitemcount}})}
\mciteSetBstSublistLabelBeginEnd{\mcitemaxwidthsubitemform\space}
{\relax}{\relax}

\bibitem[Juliano(2002)]{Juliano2002Apr}
R.~L. Juliano, \emph{Annu. Rev. Pharmacol. Toxicol.}, 2002, \textbf{42},
  283--323\relax
\mciteBstWouldAddEndPuncttrue
\mciteSetBstMidEndSepPunct{\mcitedefaultmidpunct}
{\mcitedefaultendpunct}{\mcitedefaultseppunct}\relax
\EndOfBibitem
\bibitem[Sackmann and Tanaka(2021)]{Sackmann2021Feb}
E.~Sackmann and M.~Tanaka, \emph{Biophys. Rev.}, 2021, \textbf{13},
  123--138\relax
\mciteBstWouldAddEndPuncttrue
\mciteSetBstMidEndSepPunct{\mcitedefaultmidpunct}
{\mcitedefaultendpunct}{\mcitedefaultseppunct}\relax
\EndOfBibitem
\bibitem[Heisenberg and
  Bella{\ifmmode\ddot{\imath}\else\"{\i}\fi}che(2013)]{Heisenberg2013May}
C.-P. Heisenberg and Y.~Bella{\ifmmode\ddot{\imath}\else\"{\i}\fi}che,
  \emph{Cell}, 2013, \textbf{153}, 948--962\relax
\mciteBstWouldAddEndPuncttrue
\mciteSetBstMidEndSepPunct{\mcitedefaultmidpunct}
{\mcitedefaultendpunct}{\mcitedefaultseppunct}\relax
\EndOfBibitem
\bibitem[Cavallaro and Christofori(2004)]{Cavallaro2004Feb}
U.~Cavallaro and G.~Christofori, \emph{Nat. Rev. Cancer}, 2004, \textbf{4},
  118--132\relax
\mciteBstWouldAddEndPuncttrue
\mciteSetBstMidEndSepPunct{\mcitedefaultmidpunct}
{\mcitedefaultendpunct}{\mcitedefaultseppunct}\relax
\EndOfBibitem
\bibitem[Gierer \emph{et~al.}(1972)Gierer, Berking, Bode, David, Flick,
  Hansmann, Schaller, and Trenkner]{Gierer1972Sep}
A.~Gierer, S.~Berking, H.~Bode, C.~N. David, K.~Flick, G.~Hansmann, H.~Schaller
  and E.~Trenkner, \emph{Nat. New Biol.}, 1972, \textbf{239}, 98--101\relax
\mciteBstWouldAddEndPuncttrue
\mciteSetBstMidEndSepPunct{\mcitedefaultmidpunct}
{\mcitedefaultendpunct}{\mcitedefaultseppunct}\relax
\EndOfBibitem
\bibitem[Technau and Holstein(1992)]{Technau1992May}
U.~Technau and T.~W. Holstein, \emph{Dev. Biol.}, 1992, \textbf{151},
  117--127\relax
\mciteBstWouldAddEndPuncttrue
\mciteSetBstMidEndSepPunct{\mcitedefaultmidpunct}
{\mcitedefaultendpunct}{\mcitedefaultseppunct}\relax
\EndOfBibitem
\bibitem[Bromley \emph{et~al.}(2001)Bromley, Burack, Johnson, Somersalo, Sims,
  Sumen, Davis, Shaw, Allen, and Dustin]{Bromley2001Apr}
S.~K. Bromley, W.~R. Burack, K.~G. Johnson, K.~Somersalo, T.~N. Sims, C.~Sumen,
  M.~M. Davis, A.~S. Shaw, P.~M. Allen and M.~L. Dustin, \emph{Annu. Rev.
  Immunol.}, 2001, \textbf{19}, 375--396\relax
\mciteBstWouldAddEndPuncttrue
\mciteSetBstMidEndSepPunct{\mcitedefaultmidpunct}
{\mcitedefaultendpunct}{\mcitedefaultseppunct}\relax
\EndOfBibitem
\bibitem[Balta \emph{et~al.}(2019)Balta, Monzel, Kleber, Beaudouin, Balta,
  Kaindl, Chen, Gao, Thiemann, Wirtz, Samstag, Tanaka, and
  Martin-Villalba]{Balta2019Nov}
G.~S.~G. Balta, C.~Monzel, S.~Kleber, J.~Beaudouin, E.~Balta, T.~Kaindl,
  S.~Chen, L.~Gao, M.~Thiemann, C.~R. Wirtz, Y.~Samstag, M.~Tanaka and
  A.~Martin-Villalba, \emph{Cell Rep.}, 2019, \textbf{29}, 2295--2306.e6\relax
\mciteBstWouldAddEndPuncttrue
\mciteSetBstMidEndSepPunct{\mcitedefaultmidpunct}
{\mcitedefaultendpunct}{\mcitedefaultseppunct}\relax
\EndOfBibitem
\bibitem[Komura and Andelman(2000)]{Komura2000Nov}
S.~Komura and D.~Andelman, \emph{Eur. Phys. J. E}, 2000, \textbf{3},
  259--271\relax
\mciteBstWouldAddEndPuncttrue
\mciteSetBstMidEndSepPunct{\mcitedefaultmidpunct}
{\mcitedefaultendpunct}{\mcitedefaultseppunct}\relax
\EndOfBibitem
\bibitem[Fr{\ifmmode\ddot{o}\else\"{o}\fi}hlich
  \emph{et~al.}(2021)Fr{\ifmmode\ddot{o}\else\"{o}\fi}hlich, Dasanna, Lansche,
  Czajor, Sanchez, Cyrklaff, Yamamoto, Craig, Schwarz, Lanzer, and
  Tanaka]{Frohlich2021Aug}
B.~Fr{\ifmmode\ddot{o}\else\"{o}\fi}hlich, A.~K. Dasanna, C.~Lansche,
  J.~Czajor, C.~P. Sanchez, M.~Cyrklaff, A.~Yamamoto, A.~Craig, U.~S. Schwarz,
  M.~Lanzer and M.~Tanaka, \emph{Biophys. J.}, 2021, \textbf{120},
  3315--3328\relax
\mciteBstWouldAddEndPuncttrue
\mciteSetBstMidEndSepPunct{\mcitedefaultmidpunct}
{\mcitedefaultendpunct}{\mcitedefaultseppunct}\relax
\EndOfBibitem
\bibitem[Bell \emph{et~al.}(1984)Bell, Dembo, and Bongrand]{Bell1984Jun}
G.~I. Bell, M.~Dembo and P.~Bongrand, \emph{Biophys. J.}, 1984, \textbf{45},
  1051\relax
\mciteBstWouldAddEndPuncttrue
\mciteSetBstMidEndSepPunct{\mcitedefaultmidpunct}
{\mcitedefaultendpunct}{\mcitedefaultseppunct}\relax
\EndOfBibitem
\bibitem[Bruinsma and Sackmann(2001)]{Bruinsma2001Aug}
R.~Bruinsma and E.~Sackmann, \emph{Comptes Rendus de
  l'Acad{\ifmmode\acute{e}\else\'{e}\fi}mie des Sciences - Series IV -
  Physics-Astrophysics}, 2001, \textbf{2}, 803--815\relax
\mciteBstWouldAddEndPuncttrue
\mciteSetBstMidEndSepPunct{\mcitedefaultmidpunct}
{\mcitedefaultendpunct}{\mcitedefaultseppunct}\relax
\EndOfBibitem
\bibitem[Helfrich(1973)]{Helfrich1973Dec}
W.~Helfrich, \emph{Zeitschrift f{\ifmmode\ddot{u}\else\"{u}\fi}r Naturforschung
  C}, 1973, \textbf{28}, 693--703\relax
\mciteBstWouldAddEndPuncttrue
\mciteSetBstMidEndSepPunct{\mcitedefaultmidpunct}
{\mcitedefaultendpunct}{\mcitedefaultseppunct}\relax
\EndOfBibitem
\bibitem[Gompper and Kroll(1996)]{Gompper1996Oct}
G.~Gompper and D.~M. Kroll, \emph{J. Phys. I}, 1996, \textbf{6},
  1305--1320\relax
\mciteBstWouldAddEndPuncttrue
\mciteSetBstMidEndSepPunct{\mcitedefaultmidpunct}
{\mcitedefaultendpunct}{\mcitedefaultseppunct}\relax
\EndOfBibitem
\bibitem[Gueguen \emph{et~al.}(2017)Gueguen, Destainville, and
  Manghi]{Gueguen2017Sep}
G.~Gueguen, N.~Destainville and M.~Manghi, \emph{Soft Matter}, 2017,
  \textbf{13}, 6100--6117\relax
\mciteBstWouldAddEndPuncttrue
\mciteSetBstMidEndSepPunct{\mcitedefaultmidpunct}
{\mcitedefaultendpunct}{\mcitedefaultseppunct}\relax
\EndOfBibitem
\bibitem[Tanaka and Sackmann(2005)]{Tanaka2005Sep}
M.~Tanaka and E.~Sackmann, \emph{Nature}, 2005, \textbf{437}, 656--663\relax
\mciteBstWouldAddEndPuncttrue
\mciteSetBstMidEndSepPunct{\mcitedefaultmidpunct}
{\mcitedefaultendpunct}{\mcitedefaultseppunct}\relax
\EndOfBibitem
\bibitem[Goennenwein \emph{et~al.}(2003)Goennenwein, Tanaka, Hu, Moroder, and
  Sackmann]{Goennenwein2003Jul}
S.~Goennenwein, M.~Tanaka, B.~Hu, L.~Moroder and E.~Sackmann, \emph{Biophys.
  J.}, 2003, \textbf{85}, 646--655\relax
\mciteBstWouldAddEndPuncttrue
\mciteSetBstMidEndSepPunct{\mcitedefaultmidpunct}
{\mcitedefaultendpunct}{\mcitedefaultseppunct}\relax
\EndOfBibitem
\bibitem[Purrucker \emph{et~al.}(2007)Purrucker,
  F{\ifmmode\ddot{o}\else\"{o}\fi}rtig, Jordan, Sackmann, and
  Tanaka]{Purrucker2007Feb}
O.~Purrucker, A.~F{\ifmmode\ddot{o}\else\"{o}\fi}rtig, R.~Jordan, E.~Sackmann
  and M.~Tanaka, \emph{Phys. Rev. Lett.}, 2007, \textbf{98}, 078102\relax
\mciteBstWouldAddEndPuncttrue
\mciteSetBstMidEndSepPunct{\mcitedefaultmidpunct}
{\mcitedefaultendpunct}{\mcitedefaultseppunct}\relax
\EndOfBibitem
\bibitem[Rossetti \emph{et~al.}(2015)Rossetti, Schneck, Fragneto, Konovalov,
  and Tanaka]{Rossetti2015Apr}
F.~F. Rossetti, E.~Schneck, G.~Fragneto, O.~V. Konovalov and M.~Tanaka,
  \emph{Langmuir}, 2015, \textbf{31}, 4473--4480\relax
\mciteBstWouldAddEndPuncttrue
\mciteSetBstMidEndSepPunct{\mcitedefaultmidpunct}
{\mcitedefaultendpunct}{\mcitedefaultseppunct}\relax
\EndOfBibitem
\bibitem[Alarc{\ifmmode\acute{o}\else\'{o}\fi}n
  \emph{et~al.}(2005)Alarc{\ifmmode\acute{o}\else\'{o}\fi}n, Pennadam, and
  Alexander]{Alarcon2005Feb}
C.~d. l.~H. Alarc{\ifmmode\acute{o}\else\'{o}\fi}n, S.~Pennadam and
  C.~Alexander, \emph{Chem. Soc. Rev.}, 2005, \textbf{34}, 276--285\relax
\mciteBstWouldAddEndPuncttrue
\mciteSetBstMidEndSepPunct{\mcitedefaultmidpunct}
{\mcitedefaultendpunct}{\mcitedefaultseppunct}\relax
\EndOfBibitem
\bibitem[Minko \emph{et~al.}(2003)Minko, M{\"u}ller, Motornov, Nitschke,
  Grundke, and Stamm]{Minko03b}
S.~Minko, M.~M{\"u}ller, M.~Motornov, M.~Nitschke, K.~Grundke and M.~Stamm,
  \emph{J. Am. Chem. Soc}, 2003, \textbf{125}, 3896--3900\relax
\mciteBstWouldAddEndPuncttrue
\mciteSetBstMidEndSepPunct{\mcitedefaultmidpunct}
{\mcitedefaultendpunct}{\mcitedefaultseppunct}\relax
\EndOfBibitem
\bibitem[Merlitz \emph{et~al.}(2009)Merlitz, He, Sommer, and
  Wu]{Merlitz2009Jan}
H.~Merlitz, G.-L. He, J.-U. Sommer and C.-X. Wu, \emph{Macromolecules}, 2009,
  \textbf{42}, 445--451\relax
\mciteBstWouldAddEndPuncttrue
\mciteSetBstMidEndSepPunct{\mcitedefaultmidpunct}
{\mcitedefaultendpunct}{\mcitedefaultseppunct}\relax
\EndOfBibitem
\bibitem[Hoy \emph{et~al.}(2010)Hoy, Zdyrko, Lupitskyy, Sheparovych, Aulich,
  Wang, Bittrich, Eichhorn, Uhlmann, Hinrichs, Müller, Stamm, Minko, and
  Luzinov]{MWN}
O.~Hoy, B.~Zdyrko, R.~Lupitskyy, R.~Sheparovych, D.~Aulich, J.~Wang,
  E.~Bittrich, K.-J. Eichhorn, P.~Uhlmann, K.~Hinrichs, M.~Müller, M.~Stamm,
  S.~Minko and I.~Luzinov, \emph{Adv. Funct. Mater.}, 2010, \textbf{20},
  2240--2247\relax
\mciteBstWouldAddEndPuncttrue
\mciteSetBstMidEndSepPunct{\mcitedefaultmidpunct}
{\mcitedefaultendpunct}{\mcitedefaultseppunct}\relax
\EndOfBibitem
\bibitem[Stuart \emph{et~al.}(2010)Stuart, Huck, Genzer, M{\"u}ller, Ober,
  Stamm, Sukhorukov, Szleifer, Tsukruk, Urban, Winnik, Zauscher, Luzinov, and
  Minko]{NatMat}
M.~A.~C. Stuart, W.~T.~S. Huck, J.~Genzer, M.~M{\"u}ller, C.~Ober, M.~Stamm,
  G.~B. Sukhorukov, I.~Szleifer, V.~V. Tsukruk, M.~Urban, F.~Winnik,
  S.~Zauscher, I.~Luzinov and S.~Minko, \emph{Nature Materials}, 2010,
  \textbf{9}, 101--113\relax
\mciteBstWouldAddEndPuncttrue
\mciteSetBstMidEndSepPunct{\mcitedefaultmidpunct}
{\mcitedefaultendpunct}{\mcitedefaultseppunct}\relax
\EndOfBibitem
\bibitem[Price \emph{et~al.}(2012)Price, Hur, Fredrickson, Frischknecht, and
  Huber]{Price2012Jan}
A.~D. Price, S.-M. Hur, G.~H. Fredrickson, A.~L. Frischknecht and D.~L. Huber,
  \emph{Macromolecules}, 2012, \textbf{45}, 510--524\relax
\mciteBstWouldAddEndPuncttrue
\mciteSetBstMidEndSepPunct{\mcitedefaultmidpunct}
{\mcitedefaultendpunct}{\mcitedefaultseppunct}\relax
\EndOfBibitem
\bibitem[L{\ifmmode\acute{e}\else\'{e}\fi}onforte
  \emph{et~al.}(2016)L{\ifmmode\acute{e}\else\'{e}\fi}onforte, Welling, and
  M{\ifmmode\ddot{u}\else\"{u}\fi}ller]{Leonforte2016Dec}
F.~L{\ifmmode\acute{e}\else\'{e}\fi}onforte, U.~Welling and
  M.~M{\ifmmode\ddot{u}\else\"{u}\fi}ller, \emph{J. Chem. Phys.}, 2016,
  \textbf{145}, 224902\relax
\mciteBstWouldAddEndPuncttrue
\mciteSetBstMidEndSepPunct{\mcitedefaultmidpunct}
{\mcitedefaultendpunct}{\mcitedefaultseppunct}\relax
\EndOfBibitem
\bibitem[Brown and Anseth(2017)]{Brown2017Oct}
T.~E. Brown and K.~S. Anseth, \emph{Chem. Soc. Rev.}, 2017, \textbf{46},
  6532--6552\relax
\mciteBstWouldAddEndPuncttrue
\mciteSetBstMidEndSepPunct{\mcitedefaultmidpunct}
{\mcitedefaultendpunct}{\mcitedefaultseppunct}\relax
\EndOfBibitem
\bibitem[Tanaka \emph{et~al.}(2020)Tanaka, Nakahata, Linke, and
  Kaufmann]{Tanaka2020Aug}
M.~Tanaka, M.~Nakahata, P.~Linke and S.~Kaufmann, \emph{Polym. J.}, 2020,
  \textbf{52}, 861--870\relax
\mciteBstWouldAddEndPuncttrue
\mciteSetBstMidEndSepPunct{\mcitedefaultmidpunct}
{\mcitedefaultendpunct}{\mcitedefaultseppunct}\relax
\EndOfBibitem
\bibitem[Rehfeldt \emph{et~al.}(2006)Rehfeldt, Steitz, Armes, von Klitzing,
  Gast, and Tanaka]{Rehfeldt2006May}
F.~Rehfeldt, R.~Steitz, S.~P. Armes, R.~von Klitzing, A.~P. Gast and M.~Tanaka,
  \emph{J. Phys. Chem. B}, 2006, \textbf{110}, 9171--9176\relax
\mciteBstWouldAddEndPuncttrue
\mciteSetBstMidEndSepPunct{\mcitedefaultmidpunct}
{\mcitedefaultendpunct}{\mcitedefaultseppunct}\relax
\EndOfBibitem
\bibitem[Jalilehvand \emph{et~al.}(2011)Jalilehvand, Amini, Parmar, and
  Kang]{Jalilehvand2011Nov}
F.~Jalilehvand, Z.~Amini, K.~Parmar and E.~Y. Kang, \emph{Dalton Trans.}, 2011,
  \textbf{40}, 12771--12778\relax
\mciteBstWouldAddEndPuncttrue
\mciteSetBstMidEndSepPunct{\mcitedefaultmidpunct}
{\mcitedefaultendpunct}{\mcitedefaultseppunct}\relax
\EndOfBibitem
\bibitem[Kaindl \emph{et~al.}(2012)Kaindl, Rieger, Kaschel, Engel, Schmaus,
  Sleeman, and Tanaka]{Kaindl2012Aug}
T.~Kaindl, H.~Rieger, L.-M. Kaschel, U.~Engel, A.~Schmaus, J.~Sleeman and
  M.~Tanaka, \emph{PLoS One}, 2012, \textbf{7}, e42991\relax
\mciteBstWouldAddEndPuncttrue
\mciteSetBstMidEndSepPunct{\mcitedefaultmidpunct}
{\mcitedefaultendpunct}{\mcitedefaultseppunct}\relax
\EndOfBibitem
\bibitem[Rieger \emph{et~al.}(2015)Rieger, Yoshikawa, Quadt, Nielsen, Sanchez,
  Salanti, Tanaka, and Lanzer]{Rieger2015Jan}
H.~Rieger, H.~Y. Yoshikawa, K.~Quadt, M.~A. Nielsen, C.~P. Sanchez, A.~Salanti,
  M.~Tanaka and M.~Lanzer, \emph{Blood}, 2015, \textbf{125}, 383--391\relax
\mciteBstWouldAddEndPuncttrue
\mciteSetBstMidEndSepPunct{\mcitedefaultmidpunct}
{\mcitedefaultendpunct}{\mcitedefaultseppunct}\relax
\EndOfBibitem
\bibitem[Helm \emph{et~al.}(1991)Helm, Knoll, and Israelachvili]{Helm1991}
C.~Helm, W.~Knoll and J.~Israelachvili, \emph{Proceedings of the National
  Academy of Sciences of the United States of America}, 1991, \textbf{88},
  8169--8173\relax
\mciteBstWouldAddEndPuncttrue
\mciteSetBstMidEndSepPunct{\mcitedefaultmidpunct}
{\mcitedefaultendpunct}{\mcitedefaultseppunct}\relax
\EndOfBibitem
\bibitem[Nardi \emph{et~al.}(1997)Nardi, Feder, Bruinsma, and
  Sackmann]{Nardi1997Feb}
J.~Nardi, T.~Feder, R.~Bruinsma and E.~Sackmann, \emph{Europhys. Lett.}, 1997,
  \textbf{37}, 371--376\relax
\mciteBstWouldAddEndPuncttrue
\mciteSetBstMidEndSepPunct{\mcitedefaultmidpunct}
{\mcitedefaultendpunct}{\mcitedefaultseppunct}\relax
\EndOfBibitem
\bibitem[Nardi \emph{et~al.}(1999)Nardi, Bruinsma, and Sackmann]{Nardi1999Jun}
J.~Nardi, R.~Bruinsma and E.~Sackmann, \emph{Phys. Rev. Lett.}, 1999,
  \textbf{82}, 5168--5171\relax
\mciteBstWouldAddEndPuncttrue
\mciteSetBstMidEndSepPunct{\mcitedefaultmidpunct}
{\mcitedefaultendpunct}{\mcitedefaultseppunct}\relax
\EndOfBibitem
\bibitem[Seifert and Lipowsky(1990)]{Seifert1990Oct}
U.~Seifert and R.~Lipowsky, \emph{Phys. Rev. A}, 1990, \textbf{42},
  4768--4771\relax
\mciteBstWouldAddEndPuncttrue
\mciteSetBstMidEndSepPunct{\mcitedefaultmidpunct}
{\mcitedefaultendpunct}{\mcitedefaultseppunct}\relax
\EndOfBibitem
\bibitem[Lipowsky and Seifert(1991)]{Lipowsky1991Sep}
R.~Lipowsky and U.~Seifert, \emph{Langmuir}, 1991, \textbf{7}, 1867--1873\relax
\mciteBstWouldAddEndPuncttrue
\mciteSetBstMidEndSepPunct{\mcitedefaultmidpunct}
{\mcitedefaultendpunct}{\mcitedefaultseppunct}\relax
\EndOfBibitem
\bibitem[Seifert(1991)]{Seifert1991Jun}
U.~Seifert, \emph{Phys. Rev. A}, 1991, \textbf{43}, 6803--6814\relax
\mciteBstWouldAddEndPuncttrue
\mciteSetBstMidEndSepPunct{\mcitedefaultmidpunct}
{\mcitedefaultendpunct}{\mcitedefaultseppunct}\relax
\EndOfBibitem
\bibitem[Kraus \emph{et~al.}(1995)Kraus, Seifert, and Lipowsky]{Kraus1995Nov}
M.~Kraus, U.~Seifert and R.~Lipowsky, \emph{Europhys. Lett.}, 1995,
  \textbf{32}, 431--436\relax
\mciteBstWouldAddEndPuncttrue
\mciteSetBstMidEndSepPunct{\mcitedefaultmidpunct}
{\mcitedefaultendpunct}{\mcitedefaultseppunct}\relax
\EndOfBibitem
\bibitem[Tordeux \emph{et~al.}(2002)Tordeux, Fournier, and
  Galatola]{Tordeux02b}
C.~Tordeux, J.~B. Fournier and P.~Galatola, \emph{Phys. Rev. E}, 2002,
  \textbf{65}, 041912\relax
\mciteBstWouldAddEndPuncttrue
\mciteSetBstMidEndSepPunct{\mcitedefaultmidpunct}
{\mcitedefaultendpunct}{\mcitedefaultseppunct}\relax
\EndOfBibitem
\bibitem[Matsuzaki \emph{et~al.}(2017)Matsuzaki, Ito, Chevyreva, Makky,
  Kaufmann, Okano, Kobayashi, Suganuma, Nakabayashi,
  Yoshikawa,\emph{et~al.}]{matsuzaki2017adsorption}
T.~Matsuzaki, H.~Ito, V.~Chevyreva, A.~Makky, S.~Kaufmann, K.~Okano,
  N.~Kobayashi, M.~Suganuma, S.~Nakabayashi, H.~Y. Yoshikawa \emph{et~al.},
  \emph{Physical Chemistry Chemical Physics}, 2017, \textbf{19},
  19937--19947\relax
\mciteBstWouldAddEndPuncttrue
\mciteSetBstMidEndSepPunct{\mcitedefaultmidpunct}
{\mcitedefaultendpunct}{\mcitedefaultseppunct}\relax
\EndOfBibitem
\bibitem[Lipowsky and Sackmann(1995)]{Lipowsky_Sackmann1995}
R.~Lipowsky and E.~Sackmann, \emph{{S}tructure and dynamics of {M}embranes},
  Elsevier Science, 1995\relax
\mciteBstWouldAddEndPuncttrue
\mciteSetBstMidEndSepPunct{\mcitedefaultmidpunct}
{\mcitedefaultendpunct}{\mcitedefaultseppunct}\relax
\EndOfBibitem
\bibitem[Kern and Puotinen(1983)]{Kern_Putoinen1983}
W.~Kern and D.~Puotinen, \emph{RCA Rev.}, 1983, \textbf{31}, 187--206\relax
\mciteBstWouldAddEndPuncttrue
\mciteSetBstMidEndSepPunct{\mcitedefaultmidpunct}
{\mcitedefaultendpunct}{\mcitedefaultseppunct}\relax
\EndOfBibitem
\bibitem[Hillebrandt and Tanaka(2001)]{Hillebrandt_Tanaka1983}
H.~Hillebrandt and M.~Tanaka, \emph{J. Phys. Chem. B}, 2001, \textbf{105},
  4270--4276\relax
\mciteBstWouldAddEndPuncttrue
\mciteSetBstMidEndSepPunct{\mcitedefaultmidpunct}
{\mcitedefaultendpunct}{\mcitedefaultseppunct}\relax
\EndOfBibitem
\bibitem[Parratt(1954)]{Parratt1954Jul}
L.~G. Parratt, \emph{Phys. Rev.}, 1954, \textbf{95}, 359--369\relax
\mciteBstWouldAddEndPuncttrue
\mciteSetBstMidEndSepPunct{\mcitedefaultmidpunct}
{\mcitedefaultendpunct}{\mcitedefaultseppunct}\relax
\EndOfBibitem
\bibitem[Nelson(2006)]{Nelson2006Apr}
A.~Nelson, \emph{J. Appl. Crystallogr.}, 2006, \textbf{39}, 273--276\relax
\mciteBstWouldAddEndPuncttrue
\mciteSetBstMidEndSepPunct{\mcitedefaultmidpunct}
{\mcitedefaultendpunct}{\mcitedefaultseppunct}\relax
\EndOfBibitem
\bibitem[Albersd{\"o}rfer \emph{et~al.}(1997)Albersd{\"o}rfer, Feder, and
  Sackmann]{albersdorfer1997adhesion}
A.~Albersd{\"o}rfer, T.~Feder and E.~Sackmann, \emph{Biophysical journal},
  1997, \textbf{73}, 245--257\relax
\mciteBstWouldAddEndPuncttrue
\mciteSetBstMidEndSepPunct{\mcitedefaultmidpunct}
{\mcitedefaultendpunct}{\mcitedefaultseppunct}\relax
\EndOfBibitem
\bibitem[Limozin and Sengupta(2009)]{limozin2009quantitative}
L.~Limozin and K.~Sengupta, \emph{ChemPhysChem}, 2009, \textbf{10},
  2752--2768\relax
\mciteBstWouldAddEndPuncttrue
\mciteSetBstMidEndSepPunct{\mcitedefaultmidpunct}
{\mcitedefaultendpunct}{\mcitedefaultseppunct}\relax
\EndOfBibitem
\bibitem[G{\ifmmode\acute{o}\else\'{o}\fi}{\ifmmode\acute{z}\else\'{z}\fi}d{\ifmmode\acute{z}\else\'{z}\fi}(2004)]{Gozdz2004Aug}
W.~T.
  G{\ifmmode\acute{o}\else\'{o}\fi}{\ifmmode\acute{z}\else\'{z}\fi}d{\ifmmode\acute{z}\else\'{z}\fi},
  \emph{Langmuir}, 2004, \textbf{20}, 7385--7391\relax
\mciteBstWouldAddEndPuncttrue
\mciteSetBstMidEndSepPunct{\mcitedefaultmidpunct}
{\mcitedefaultendpunct}{\mcitedefaultseppunct}\relax
\EndOfBibitem
\bibitem[Raval and
  G{\ifmmode\acute{o}\else\'{o}\fi}{\ifmmode\acute{z}\else\'{z}\fi}d{\ifmmode\acute{z}\else\'{z}\fi}(2020)]{Raval2020Jul}
J.~Raval and W.~T.
  G{\ifmmode\acute{o}\else\'{o}\fi}{\ifmmode\acute{z}\else\'{z}\fi}d{\ifmmode\acute{z}\else\'{z}\fi},
  \emph{ACS Omega}, 2020, \textbf{5}, 16099--16105\relax
\mciteBstWouldAddEndPuncttrue
\mciteSetBstMidEndSepPunct{\mcitedefaultmidpunct}
{\mcitedefaultendpunct}{\mcitedefaultseppunct}\relax
\EndOfBibitem
\bibitem[Helfrich(1973)]{Helfrich1973}
W.~Helfrich, \emph{Z. Naturforsch.}, 1973, \textbf{28c}, 693--703\relax
\mciteBstWouldAddEndPuncttrue
\mciteSetBstMidEndSepPunct{\mcitedefaultmidpunct}
{\mcitedefaultendpunct}{\mcitedefaultseppunct}\relax
\EndOfBibitem
\bibitem[Nardi \emph{et~al.}(1998)Nardi, Bruinsma, and
  Sackmann]{nardi1998adhesion}
J.~Nardi, R.~Bruinsma and E.~Sackmann, \emph{Physical Review E}, 1998,
  \textbf{58}, 6340\relax
\mciteBstWouldAddEndPuncttrue
\mciteSetBstMidEndSepPunct{\mcitedefaultmidpunct}
{\mcitedefaultendpunct}{\mcitedefaultseppunct}\relax
\EndOfBibitem
\bibitem[Nistor \emph{et~al.}(2014)Nistor, Pamfil, Schick, and
  Vasile]{NISTOR2014114}
M.-T. Nistor, D.~Pamfil, C.~Schick and C.~Vasile, \emph{Thermochimica Acta},
  2014, \textbf{589}, 114--122\relax
\mciteBstWouldAddEndPuncttrue
\mciteSetBstMidEndSepPunct{\mcitedefaultmidpunct}
{\mcitedefaultendpunct}{\mcitedefaultseppunct}\relax
\EndOfBibitem
\bibitem[Bruinsma(1995)]{bruinsma1995physics}
R.~Bruinsma, \emph{Physics of Biomaterials, Physical aspects of adhesion of
  leukocytes}, Kluwer, Dordrecht,, 1995, pp. 61--102\relax
\mciteBstWouldAddEndPuncttrue
\mciteSetBstMidEndSepPunct{\mcitedefaultmidpunct}
{\mcitedefaultendpunct}{\mcitedefaultseppunct}\relax
\EndOfBibitem
\bibitem[Schmidt \emph{et~al.}(2014)Schmidt, Monzel, Bihr, Merkel, Seifert,
  Sengupta, and Smith]{Schmidt2014May}
D.~Schmidt, C.~Monzel, T.~Bihr, R.~Merkel, U.~Seifert, K.~Sengupta and A.-S.
  Smith, \emph{Phys. Rev. X}, 2014, \textbf{4}, 021023\relax
\mciteBstWouldAddEndPuncttrue
\mciteSetBstMidEndSepPunct{\mcitedefaultmidpunct}
{\mcitedefaultendpunct}{\mcitedefaultseppunct}\relax
\EndOfBibitem
\bibitem[Derjaguin \emph{et~al.}(1987)Derjaguin, Churaev, Muller, and
  Kisin]{derjaguin1987surface}
B.~V. Derjaguin, N.~V. Churaev, V.~M. Muller and V.~Kisin, \emph{Surface
  forces}, Springer, 1987\relax
\mciteBstWouldAddEndPuncttrue
\mciteSetBstMidEndSepPunct{\mcitedefaultmidpunct}
{\mcitedefaultendpunct}{\mcitedefaultseppunct}\relax
\EndOfBibitem
\bibitem[Swain and Andelman(1999)]{swain1999influence}
P.~S. Swain and D.~Andelman, \emph{Langmuir}, 1999, \textbf{15},
  8902--8914\relax
\mciteBstWouldAddEndPuncttrue
\mciteSetBstMidEndSepPunct{\mcitedefaultmidpunct}
{\mcitedefaultendpunct}{\mcitedefaultseppunct}\relax
\EndOfBibitem
\bibitem[Tanaka(2013)]{tanaka2013physics}
M.~Tanaka, \emph{Current opinion in colloid \& interface science}, 2013,
  \textbf{18}, 432--439\relax
\mciteBstWouldAddEndPuncttrue
\mciteSetBstMidEndSepPunct{\mcitedefaultmidpunct}
{\mcitedefaultendpunct}{\mcitedefaultseppunct}\relax
\EndOfBibitem
\bibitem[Helfrich and Servuss(1984)]{Helfrich84}
W.~Helfrich and R.~M. Servuss, \emph{Nuovo Cimento Della Societa Italiana Di
  Fisica D-Condensed Matter}, 1984, \textbf{3}, 137--151\relax
\mciteBstWouldAddEndPuncttrue
\mciteSetBstMidEndSepPunct{\mcitedefaultmidpunct}
{\mcitedefaultendpunct}{\mcitedefaultseppunct}\relax
\EndOfBibitem
\bibitem[Derjaguin(1934)]{Derjaguin1934Nov}
B.~Derjaguin, \emph{Kolloid-Zeitschrift}, 1934, \textbf{69}, 155--164\relax
\mciteBstWouldAddEndPuncttrue
\mciteSetBstMidEndSepPunct{\mcitedefaultmidpunct}
{\mcitedefaultendpunct}{\mcitedefaultseppunct}\relax
\EndOfBibitem
\bibitem[Evans(1990)]{Evans1990Jan}
E.~Evans, \emph{Colloids Surf.}, 1990, \textbf{43}, 327--347\relax
\mciteBstWouldAddEndPuncttrue
\mciteSetBstMidEndSepPunct{\mcitedefaultmidpunct}
{\mcitedefaultendpunct}{\mcitedefaultseppunct}\relax
\EndOfBibitem
\bibitem[Guttenberg \emph{et~al.}(2000)Guttenberg, Bausch, Hu, Bruinsma,
  Moroder, and Sackmann]{Guttenberg2000Nov}
Z.~Guttenberg, A.~R. Bausch, B.~Hu, R.~Bruinsma, L.~Moroder and E.~Sackmann,
  \emph{Langmuir}, 2000, \textbf{16}, 8984--8993\relax
\mciteBstWouldAddEndPuncttrue
\mciteSetBstMidEndSepPunct{\mcitedefaultmidpunct}
{\mcitedefaultendpunct}{\mcitedefaultseppunct}\relax
\EndOfBibitem
\bibitem[Schick(1990)]{schick90}
M.~Schick, \emph{Les Houches lectures on ``Liquids at Interfaces''}, Elsevier
  Science Publishers, Amsterdam, 1990, vol. Session XLVIII, pp. 1--89\relax
\mciteBstWouldAddEndPuncttrue
\mciteSetBstMidEndSepPunct{\mcitedefaultmidpunct}
{\mcitedefaultendpunct}{\mcitedefaultseppunct}\relax
\EndOfBibitem
\bibitem[{M. M{\"u}ller} and {L. G. Mac{D}owell}(2003)]{Muller2003}
{M. M{\"u}ller} and {L. G. Mac{D}owell}, \emph{J. Phys. Cond. Matt.}, 2003,
  \textbf{15}, R609--R653\relax
\mciteBstWouldAddEndPuncttrue
\mciteSetBstMidEndSepPunct{\mcitedefaultmidpunct}
{\mcitedefaultendpunct}{\mcitedefaultseppunct}\relax
\EndOfBibitem
\bibitem[Indekeu \emph{et~al.}(1999)Indekeu, Ragil, Bonn, Broseta, and
  Meunier]{Indekeu1999Jun}
J.~O. Indekeu, K.~Ragil, D.~Bonn, D.~Broseta and J.~Meunier, \emph{J. Stat.
  Phys.}, 1999, \textbf{95}, 1009--1043\relax
\mciteBstWouldAddEndPuncttrue
\mciteSetBstMidEndSepPunct{\mcitedefaultmidpunct}
{\mcitedefaultendpunct}{\mcitedefaultseppunct}\relax
\EndOfBibitem
\bibitem[Binder \emph{et~al.}(2003)Binder, Landau, and M{\"u}ller]{Binder03}
K.~Binder, D.~Landau and M.~M{\"u}ller, \emph{J. Stat. Phys.}, 2003,
  \textbf{110}, 1411--1514\relax
\mciteBstWouldAddEndPuncttrue
\mciteSetBstMidEndSepPunct{\mcitedefaultmidpunct}
{\mcitedefaultendpunct}{\mcitedefaultseppunct}\relax
\EndOfBibitem
\bibitem[Bonn \emph{et~al.}(2009)Bonn, Eggers, Indekeu, Meunier, and
  Rolley]{Bonn2009May}
D.~Bonn, J.~Eggers, J.~Indekeu, J.~Meunier and E.~Rolley, \emph{Rev. Mod.
  Phys.}, 2009, \textbf{81}, 739--805\relax
\mciteBstWouldAddEndPuncttrue
\mciteSetBstMidEndSepPunct{\mcitedefaultmidpunct}
{\mcitedefaultendpunct}{\mcitedefaultseppunct}\relax
\EndOfBibitem
\bibitem[Helfrich(1978)]{Helfrich1978Mar}
W.~Helfrich, \emph{Zeitschrift f{\ifmmode\ddot{u}\else\"{u}\fi}r Naturforschung
  A}, 1978, \textbf{33}, 305--315\relax
\mciteBstWouldAddEndPuncttrue
\mciteSetBstMidEndSepPunct{\mcitedefaultmidpunct}
{\mcitedefaultendpunct}{\mcitedefaultseppunct}\relax
\EndOfBibitem
\bibitem[Peliti and Leibler(1985)]{Peliti1985Apr}
L.~Peliti and S.~Leibler, \emph{Phys. Rev. Lett.}, 1985, \textbf{54},
  1690--1693\relax
\mciteBstWouldAddEndPuncttrue
\mciteSetBstMidEndSepPunct{\mcitedefaultmidpunct}
{\mcitedefaultendpunct}{\mcitedefaultseppunct}\relax
\EndOfBibitem
\bibitem[Gruhn and Lipowsky(2005)]{Gruhn2005Jan}
T.~Gruhn and R.~Lipowsky, \emph{Phys. Rev. E}, 2005, \textbf{71}, 011903\relax
\mciteBstWouldAddEndPuncttrue
\mciteSetBstMidEndSepPunct{\mcitedefaultmidpunct}
{\mcitedefaultendpunct}{\mcitedefaultseppunct}\relax
\EndOfBibitem
\end{mcitethebibliography}

\providecommand*{\mcitethebibliography}{\thebibliography}
\csname @ifundefined\endcsname{endmcitethebibliography}
{\let\endmcitethebibliography\endthebibliography}{}

\end{document}